\documentclass{article}
\usepackage{bookstyle,bm,cite,latexsym}
\usepackage[dvips]{graphicx}
\usepackage{color,amsmath,amssymb} 
\usepackage{psfrag}

\def\laq{\raise 0.4ex\hbox{$<$}\kern -0.8em\lower 0.62ex\hbox{$\sim$}}
\def\gaq{\raise 0.4ex\hbox{$>$}\kern -0.7em\lower 0.62ex\hbox{$\sim$}}

\newcommand{\beq}{\begin{equation}}
\newcommand{\eeq}{\end{equation}}
\newcommand{\bea}{\begin{eqnarray}} 
\newcommand{\eea}{\end{eqnarray}}
\newcommand{\ba}{\begin{array}}
\newcommand{\ea}{\end{array}}

\newcommand{\mytextrm}[1]{{}}

\newcommand{\hS}{\mbox{\boldmath${\widehat{S}}$}}

\newcommand{\vV}{\mbox{\boldmath${V}$}}

\newcommand{\vS}{\mbox{\boldmath${S}$}}

\newcommand{\vX}{\mbox{\boldmath${X}$}}

\newlength{\sizeonefig}
\newlength{\sizetwofig}
\newlength{\sizeonefigb}
\newlength{\sizetwofigb}
\begin{document}
\title{Gravitational waves}
\author{Alessandra Buonanno}
\address{Department of Physics, University of Maryland,\\ College Park MD 20742, USA~
\footnote{Lectures given at the Fabric of Spacetime Summer School (2006), Les Houches, France.}\\
{\tt March 31, 2007}}
\frontmatter
\maketitle
\mainmatter%

\section{Introduction}
\label{sec1}

Gravitational-wave (GW) science has entered a new era. 
Experimentally~\footnote{GW experiments started with the pioneering work of Joseph Weber 
at Maryland in the 60s}, several ground-based laser-interferometer 
GW detectors ($ 10 \mbox{--} 1$ kHz) have been built 
in the United States (LIGO)~\cite{LIGO}, Europe (VIRGO and GEO)
~\cite{VIRGO,GEO} and Japan (TAMA)~\cite{TAMA}, and are now taking data at 
design sensitivity. Advanced optical configurations capable 
of reaching sensitivities slightly above and
even below the so-called standard-quantum-limit 
for a {\it free} test-particle, have been designed for second~\cite{AdIFO2} and 
third generation~\cite{AdIFO3} GW detectors ($\sim 2011\mbox{--}2020$). A laser 
interferometer space antenna (LISA)~\cite{LISA} 
($10^{-4} \mbox{--} 10^{-2}$ Hz) might fly within  
the next decade. Resonant-bar detectors ($\sim 1$ kHz)~\cite{bars}  
are improving more and more their sensitivity, 
broadening their frequency band. At much lower frequencies, 
$\sim 10^{-17}$ Hz, future cosmic microwave background (CMB) probes might 
{\it detect} GWs by measuring the CMB polarization~\cite{pol}. 
Millisecond pulsar timing can set interesting 
upper limits~\cite{jenetetal}  in the frequency range $10^{-9} \mbox{--} 
10^{-8}$\,Hz. At such frequencies, the large number of millisecond pulsars 
which will be detectable with the square kilometer array~\cite{SKAref}, 
would provide an ensemble of clocks that can be used as multiple arms 
of a GW detector.

Theoretically, the last years have been characterized by 
numerous major advances. For what concerns the most promising 
GW sources for ground-based and space-based detectors, notably, 
binary systems composed of neutron stars (NS) 
and black holes (BHs), our understanding of the two-body problem 
and the GW-generation problem has improved significantly. 
The best-developed {\it analytic} approximation method 
in general relativity is undoubtably the post-Newtonian (PN) 
method. The errors and ambiguities 
that characterized the very early literature on the PN problem 
of motion (for a review see, e.g., Ref.\cite{TD87}), 
have been overcome. Robust predictions 
are currently available through 3.5PN order~\cite{PNnospin} ($v^7/c^7$),  
if the compact objects do not carry spin, and 2.5PN order~\cite{PNspin} 
($v^5/c^5$) if they carry spin. Resummation of the PN expansion aimed 
at pushing analytic calculations until the final stage of evolution, 
including the transition inspiral--merger--ring-down,   
have also been proposed, for the conservative two-body dynamics~\cite{EOB} 
and the radiation-reaction effects~\cite{DIS98}. Quite interestingly, 
the effective-field-theory approach, commonly 
used in particle physics, has been extended to gravity, 
notably to the two-body problem of motion~\cite{EFT}. 
The recent dazzling  breakthrough in {\it numerical relativity}~\cite{NR}, with 
different independent groups being able to successfully 
evolve a comparable-mass BH binary throughout inspiral, merger and ring-down 
and extract the GW signal, is allowing to dig out details of 
the nonlinear dynamics which could not be {\it fully} 
predicted with other means. 

Our knowledge has also progressed on the problem 
of motion of a point particle in curved spacetime 
when the emission of GWs is taken into account 
(nongeodesic motion)~\cite{RR,others}. To solve this problem 
is of considerable importance for predicting very accurate waveforms 
emitted by extreme mass-ratio binaries, which are among the most promising 
sources for LISA~\cite{emri}.  

The GW community working at the interface between 
the theory and the experiment has provided 
{\it templates}~\cite{templates,DIS98,EOB} for binaries 
and developed robust algorithms~\cite{DA,algorithms} for 
pulsars and other GW sources observable 
with ground-based and space-based interferometers. The joined work of 
data analysts and experimentalists has established 
astrophysically significant upper limits for several GW sources
~\cite{lsc,lscpulsar,lscstoch} and  
is now eagerly waiting for the first detection. 

These lectures were envisioned to be an introductory, basic 
course in GW theory. Many of the topics that we shall address 
are thoroughly discussed in several books~\cite{LL,MTW,SW,BW,BS,SC,M07} and 
proceedings or reviews~\cite{KT87,BA,M00,CT02,AB03,FH05,KTcaltech}. The lectures focused more 
on binary systems, probably because biased towards' the author own background 
and expertise. The lectures are organized as follows. 
In Sec.~\ref{sec2} we start by deriving the wave equation in linearized 
gravity and discuss the main properties of GWs. 
In Sec.~\ref{sec3} we describe the interaction of GWs with 
free test particles and the key ideas underlying the functioning 
of GW detectors. Section~\ref{sec4} reviews the effective 
stress-energy tensor of GWs. Section~\ref{sec5} is devoted 
to the generation problem. We explicitly derive the gravitational field 
at leading order assuming slow-motion, weak-gravity  
and negligible self-gravity. We then discuss how 
those results can be extended to non-negligible self-gravity 
sources. As a first application, in Sec.~\ref{sec6} we 
compute the GW signal from binary systems. We discuss briefly 
the state-of-the-art of PN calculations and NR results. 
As an example of data-analysis issues, we compute 
the GW templates in the stationary-phase-approximation (SPA). 
In Sec.~\ref{sec7} we apply the results of Sec.~\ref{sec6} 
to other astrophysical sources, notably pulsars 
and supernovae. Section~\ref{sec8} 
focus on cosmological sources at much higher red-shift 
$z \gg 1$. We review the main physical mechanisms 
that could have produced GWs in the early Universe, notably 
first-order phase transitions, cosmic and fundamental strings,  
and inflation. 

\section{Linearization of Einstein equations}
\label{sec2}

In 1916 Einstein realized the propagation effects at finite 
velocity in the gravitational equations and predicted the 
existence of wave-like solutions of the linearized vacuum 
field equations~\cite{E16}. In this section we expand  
Einstein equations around the flat Minkowski metric 
derive the wave equation and put the solution in a 
simple form using an appropriate gauge.

\subsection{Einstein equations and gauge symmetry}
\label{sec2.1}

The Einstein action reads
\beq
S_{\rm g} = \frac{c^3}{16 \pi\,G}\,\int d^4 x \,\sqrt{-g}\, {R}\,,
\label{eq:1}
\eeq
where $c$ denotes the speed of light, $G$ the Newton constant, 
$g_{\mu \nu}$ is the four dimensional metric and $g =\det (g_{\mu \nu})$. 
Henceforth, we use the following conventions. The flat Minkowski 
metric is $\eta_{\mu \nu} = (-,+,+,+)$, Greek indices denote 
spacetime coordinates $\mu, \nu = 0,1,2,3$, whereas Latin indices 
denote spacelike coordinates $i,j = 1,2,3$. 
Moreover, $x^\mu = (x^0,\mathbf{x}) = (c\,t,\mathbf{x})$, thus 
$d^4 x = c\, dt\,d^3x$. Partial derivatives 
$\partial_\mu$ will be denoted with a comma, while 
covariant derivatives with a semicolon. 
The scalar tensor ${R}$ in Eq.~(\ref{eq:1}) is 
obtained from the curvature tensor as
\bea
&& {R} = g^{\mu \nu}\,{R}_{\mu \nu}\,, \quad \quad 
{R}_{\mu \nu} = g^{\rho \sigma}\,R_{\rho \mu \sigma \nu}\,,\\
&& R^\nu_{\,\,\,\,\mu \rho \sigma} = 
\frac{\partial \Gamma_{\mu \sigma}^\nu}{\partial x^\rho} - 
\frac{\partial \Gamma_{\mu \rho}^\nu}{\partial x^\sigma} + 
\Gamma^\nu_{\lambda \rho}\,\Gamma_{\mu \sigma}^\lambda - 
\Gamma^\nu_{\lambda \sigma}\,\Gamma_{\mu \rho}^\lambda \,,
\label{eq:2}
\eea
where $\Gamma^\nu_{\mu \sigma}$ are the affine connections 
\beq 
\Gamma^\mu_{\nu \rho} = \frac{1}{2}\,g^{\mu \lambda}\, 
\left (\frac{\partial g_{\lambda \nu}}{\partial x^\rho} +
\frac{\partial g_{\lambda \rho}}{\partial x^\nu} -
\frac{\partial g_{\nu \rho}}{\partial x^\lambda} \right )\,.
\label{eq:3}
\eeq
The curvature tensor satisfies the following properties
\bea
&& R_{\mu \nu \rho \sigma} = - R_{\nu \mu \rho \sigma} = - R_{\mu \nu \sigma \rho}\,, \quad  
R_{\mu \nu \rho \sigma} = R_{\rho \sigma \mu \nu}\,,\\
&& 
R_{\mu \nu \rho \sigma} + R_{\mu \sigma \nu \rho} + R_{\mu \rho \sigma \nu} =0\,, 
\quad  R^\lambda_{\;\;\mu \nu \rho; \sigma} + 
R^\lambda_{\;\;\mu \sigma \nu; \rho} + R^\lambda_{\;\;\mu \rho \sigma; \nu} =0\,. \nonumber \\
\eea
The latter equation is known as the Bianchi identity. 
We define the matter energy-momentum tensor $T_{\mu \nu}$ from the 
variation of the matter action $S_{\rm m}$ under a change of 
the metric $g_{\mu \nu} \rightarrow g_{\mu \nu} + \delta g_{\mu \nu}$, that is  
\beq
\delta S_{\rm m} = \frac{1}{2c}\,\int d^4 x\,\sqrt{-g}\,T^{\mu \nu}\,
\delta g_{\mu \nu}\,.
\label{eq:4}
\eeq
The variation of the total action $S= S_{\rm g}+S_{\rm m} $ with respect to 
$g_{\mu \nu}$ gives the Einstein equations
\beq
G_{\mu \nu} = {R}_{\mu \nu} - \frac{1}{2}\,g_{\mu \nu}\,{R} = 
\frac{8 \pi\,G}{c^4}\,T_{\mu \nu}\,.
\label{eq:5}
\eeq 
The above equations are nonlinear equations with well posed initial 
value structure, i.e. they determine future values of $g_{\mu \nu}$ 
from given initial values. Since $\mu =0,\cdots 3, \nu=0, \cdots 3$, 
Eq.~(\ref{eq:5}) contains sixteen differential equations, which reduce 
to ten differential equations if the symmetry of the tensors 
$G_{\mu \nu}$ and $T_{\mu \nu}$ is used. Finally, because of 
the Bianchi identity we have $G_{\mu \nu}^{\;\;\;\; ;\nu} = 0$, thus 
the ten differential equations reduce to six.

General relativity is invariant under the group of all possible coordinate 
transformations 
\beq
x^\mu \rightarrow x'^{\mu}(x)\,,
\label{eq:6}
\eeq
where $x'^{\mu}$ is invertible, differentiable and with a differentiable 
inverse. Under the above transformation, the metric transforms as 
\beq
g_{\mu \nu}(x) \rightarrow g'_{\mu \nu}(x') = \frac{\partial x^\rho}{\partial x'^\mu}\,
\frac{\partial x^\sigma}{\partial x'^\nu}\,g_{\rho \sigma}(x)\,.
\label{eq:7}
\eeq
We assume that there exists a reference frame in which, on a sufficiently 
large spacetime region, we can write 
\beq
g_{\mu \nu} = \eta_{\mu \nu} + h_{\mu \nu}\,, \quad \quad |h_{\mu \nu}| \ll 1\,.
\label{eq:7b}
\eeq
By choosing this particular reference frame, we break the invariance of general 
relativity under coordinate transformations. However, a residual gauge 
symmetry remains. Let us consider the following coordinate transformation 
\beq 
x^\mu \rightarrow x'^\mu = x^\mu + \xi^\mu(x)\,, \quad \quad |\partial_\mu \xi_\nu| 
\leq |h_{\mu \nu}|\,. 
\label{eq:8}
\eeq
The metric transforms as 
\beq 
g'_{\mu \nu} = \eta_{\mu \nu} - \partial_\nu \xi_\mu - \partial_\mu \xi_\nu 
+ h_{\mu \nu} + {\cal O}(\partial \xi^2)\,,
\label{eq:9}
\eeq
thus, introducing 
\beq 
h'_{\mu \nu} = h_{\mu \nu} - \xi_{\mu,\nu} - \xi_{\nu,\mu}\,,
\label{eq:10}
\eeq
we have 
\beq
g'_{\mu \nu} = \eta_{\mu \nu} + h'_{\mu \nu}\,, \quad \quad |h'_{\mu \nu}| \ll 1\,.
\label{eq:11}
\eeq
In conclusion, the slowly varying coordinate transformations (\ref{eq:8}) 
are a symmetry of the linearized theory. 
Under a finite, global ($x$-independent) Lorentz transformation 
\beq
x^\mu \rightarrow \Lambda_\nu^\mu\, x^\nu\,, \quad \quad  \Lambda_\mu^\rho\,\Lambda_\nu^\sigma\,
\eta_{\rho \sigma} = \eta_{\mu \nu}\,,
\label{eq:12}
\eeq
the metric transforms as 
\beq
g_{\mu \nu} \rightarrow g'_{\mu \nu}(x') = \Lambda_\mu^\rho\,
\Lambda_\nu^\sigma\,g_{\rho \sigma} = \eta_{\mu \nu} + \Lambda_\mu^\rho\,
\Lambda_\nu^\sigma\,h_{\rho \sigma}(x) = \eta_{\mu \nu} + h'_{\mu \nu}(x')\,,
\label{eq:13}
\eeq
thus, $h_{\mu \nu}$ is a tensor under Lorentz transformations. It is straightforward 
to prove that $h_{\mu \nu}$ is also invariant under translations. In conclusions, linearized 
theory is invariant under the Poincar\'e group and under the transformation 
$x^\mu \rightarrow x^\mu + \xi^\mu$ with $|\partial_\nu \xi^\mu | \ll 1 $.

\subsection{Wave equation}
\label{sec2.2}

Let us now linearize Einstein equations posing 
$g_{\mu \nu} = \eta_{\mu \nu} + h_{\mu \nu}$.  At linear order in $h_{\mu \nu}$ 
the affine connections and curvature tensor read 
\bea
&& \Gamma_{\mu \rho}^\nu = \frac{1}{2}\eta^{\nu \lambda}\, 
(\partial_\rho h_{\lambda \mu} + \partial_\mu h_{\lambda \rho} 
- \partial_\lambda h_{\mu \rho} )\,,\\
&& {R}^\nu_{\mu \rho \sigma} = \partial_\rho \Gamma_{\mu \sigma}^\nu 
- \partial_\sigma \Gamma_{\mu \rho}^\nu + {\cal O}(h^2)\,,
\label{eq:14}
\eea
more explicitly
\beq
R_{\mu \nu \rho \sigma} = \frac{1}{2}\,\left ( \partial_{\rho \nu} 
h_{\mu \sigma} + \partial_{\sigma \mu} h_{\nu \rho} - 
\partial_{\rho \mu} h_{\nu \sigma} - \partial_{\sigma \nu} h_{\mu \rho}\right )\,. 
\label{eq:15}
\eeq
Using the above equations, it is straightforward to show that 
the linearized Riemann tensor is invariant 
under the transformation $h_{\mu \nu} \rightarrow h_{\mu \nu} - \partial_\mu \xi_\nu  
- \partial_\nu \xi_\mu $. Equation (\ref{eq:15}) can be greatly simplified 
if we introduce the so-called {\it trace-reverse} tensor 
\beq
\overline{h}^{\mu \nu} = h^{\mu \nu} - \frac{1}{2}\eta^{\mu \nu}\,h\,, 
\label{eq:16}
\eeq
where $h = \eta_{\alpha \beta}\,h^{\alpha \beta}$ and $\overline{h} = -h$, which 
explains the name. Some algebra leads to 
\beq
\Box \overline{h}_{\nu \sigma} + 
\eta_{\nu \sigma}\,\partial^\rho\,\partial^\lambda \overline{h}_{\rho \lambda} 
- \partial^\rho\,\partial_\nu \overline{h}_{\rho \sigma} 
- \partial^\rho\,\partial_\sigma \overline{h}_{\rho \nu}  + 
{\cal O}(h^2) = - \frac{16 \pi G}{c^4}\,T_{\nu \sigma}\,,
\label{eq:17}
\eeq
where the wave operator $\Box = \eta_{\rho \sigma}\,\partial^\rho\,\partial^\sigma$. 
To further simplify Eq.~(\ref{eq:17}) we can impose the Lorenz gauge 
(also denoted in the literature as harmonic or De Donder gauge)
\beq
\partial_\nu \overline{h}^{\mu \nu} = 0\,,
\label{eq:18}
\eeq
and obtain
\beq
\Box \overline{h}_{\nu \sigma} = 
- \frac{16 \pi G}{c^4}\,T_{\nu \sigma}\,.
\label{eq:19}
\eeq
If $\overline{h}^{\mu \nu}$ does not satisfy the Lorenz gauge, 
i.e. $\partial_\mu \overline{h}^{\mu \nu} = q_{\nu}$,  we 
can introduce a coordinate transformation such that 
$\overline{h'}_{\mu \nu} = \overline{h}_{\mu \nu} - \xi_{\mu ,\nu} 
- \xi_{\nu ,\mu} + \eta_{\mu \nu}(\partial_\rho \xi^\rho)$ 
and impose $\Box \xi_\nu = q_\nu$, 
obtaining $\partial_\mu \overline{h'}^{\mu \nu} = 0$.

Summarizing, the Lorenz gauge imposes 4 conditions 
that allow to reduce the 10 independent components 
of the $4 \times 4$ symmetric tensor $h_{\mu \nu}$ to 6  
independent components. Note that we also have the 
condition $\partial_\mu T^{\mu \nu} = 0$, which is the 
conservation of the energy-momentum tensor 
of the matter in linearized theory. By contrast in the 
full theory $T^{\mu \nu}_{\;\;\;; \nu} = 0$. 

\subsection{Transverse-traceless gauge}
\label{sec2.3}

We want to study the propagation of GWs once they have 
been generated. We set $T_{\mu \nu} = 0$ in Eq.~(\ref{eq:19}) 
and obtain the wave equation in vacuum
\beq 
\Box \overline{h}_{\mu \nu} =0\,.
\label{eq:20}
\eeq
GWs propagate at the speed of light. 
Within the Lorenz gauge we can always consider coordinate 
transformations such that $\Box \xi_\mu =0$. The trace-reverse tensor 
transforms as $\overline{h'}_{\mu \nu} = \overline{h}_{\mu \nu} + 
\xi_{\mu \nu}$ with $\xi_{\mu \nu} = 
\eta_{\mu \nu}\,\partial_\rho \xi^\rho - \xi_{\mu,\nu} - \xi_{\nu,\mu}$. 
Using $\Box \xi_{\mu \nu}=0$, we can subtract 4 of 
the 6 components of $\overline{h}_{\mu \nu}$. More specifically,   
we can choose $\xi^0$ such that $\overline{h} = 0$ and $\xi^i$ 
such that $h^{i 0}=0$, thus $\partial_0 h^{00}=0$. 
The GW being a time-dependent field, we can set $h^{00}=0$. 
We denote the field $h_{i j}$ which satisfies the following 
transverse and traceless gauge conditions, 
\beq 
h^{00} = 0\,, \quad h^{0 i} =0\,, \quad \partial_i h^{i j} = 0\,,
\quad h^{ii} = 0\,,
\label{eq:21}
\eeq
the {\it transverse-traceless} tensor $h_{i j}^{\rm TT}$. Note that 
for a single plane wave with wave vector $\mathbf{k}$ and 
propagation direction $\mathbf{n} = \mathbf{k}/k$, the 
transversality condition reduces to $n^i\,h_{i j}^{\rm TT}=0$. 
Without loosing in generality, we can assume that the plane 
wave propagates along the $z$-axis, thus
\beq
h_{i j}^{\rm TT}(t,z) = 
\left (\begin{array}{ccc}
h_+ & h_\times & 0\\
h_\times & - h_+ & 0 \\
0  & 0 & 0 
\end{array}
\right )
\,\cos \left [ \omega\,\left (t - \frac{z}{c}\right )\right ]\,,
\label{eq:22}
\eeq
where we indicate with $h_+$ and $h_\times$ the two independent 
polarization states. Following~\cite{SW,M07}, we can introduce 
the projector operator $P_{i j}(\mathbf{n}) = \delta_{i j} - n_i\,n_j$, 
which satisfies the conditions 
\beq
P_{i j} = P_{j i}\,, \quad n^i\,P_{i j} = 0\,, \quad 
P_{i j}\,P^{j k} = P_i^k \,,\quad  P_{ii} =2\,,
\label{eq:23}
\eeq
and the $\Lambda$-operator 
\beq
\Lambda_{ij\,kl}(\mathbf{n}) = P_{i k}\,P_{j l} - \frac{1}{2}\,
P_{i j}\,P_{kl}\,,
\label{eq:24}
\eeq
and obtain the TT field for a generic propagation direction 
\beq
h_{i j}^{\rm TT} = \Lambda_{i j, kl}\,h_{k l}\,,
\label{eq:25}
\eeq
where $h_{k l}$ is given in the Lorenz gauge but not necessarily 
in the TT gauge.

The GW is described in the TT gauge by a $2 \times 2$ matrix in 
the plane orthogonal to the direction of propagation $\mathbf{n}$. 
If we perform a rotation $\psi$ about the axis $\mathbf{n}$, we obtain 
\beq 
h_\times \pm i\,h_+ \rightarrow e^{\mp 2 i\,\psi}\,(h_\times \pm i\,h_+)\,.
\label{eq:26}
\eeq
In particle physics we call helicity the projection of 
the (total) angular momentum 
along the propagation direction: ${\cal H} = \mathbf{J} \cdot \mathbf{n} = 
\mathbf{S} \cdot \mathbf{n} $, being $\mathbf{S}$ the particle's spin. 
Under a rotation $\psi$ about the propagation direction 
the helicity states transform as $ h \rightarrow e^{i {\cal H} \,\psi}\,h$. 
Thus, Eq.~(\ref{eq:26}) states that $h_{\times}- ih_+$ are the 
helicity states and that the graviton is a spin-2 particle.

\section{Interaction of gravitational waves with point particles}
\label{sec3}

\subsection{Newtonian and relativistic description of tidal gravity}

Let us consider two point particles, labeled A and B, falling freely 
through 3-D Euclidean space under the action of an external Newtonian potential 
$\Phi$. We assume that at time $t=0$ the particles are separated 
by a small distance $\mathbf{\xi}$ and have initially the same speed 
$\mathbf{v}_{\rm A}(t=0)=\mathbf{v}_{\rm B}(t=0)$. Since the two 
particles are at slightly different positions, they experience a slightly 
different gravitational potential $\Phi$ and a 
different acceleration $\mathbf{g} = - \boldsymbol{\nabla} \Phi$. At time 
$t> 0$, $\mathbf{v}_{\rm A}(t) \neq \mathbf{v}_{\rm B}(t)$. 
Let us introduce the separation vector $\boldsymbol{\xi} = \mathbf{x}_A - 
\mathbf{x}_B$ in 3-D Euclidean space. We have 
\beq
\frac{d^2 \xi^i}{d t^2} = - \left (\frac{\partial \Phi}{d x^i}\right )_{\rm B}  
+  \left (\frac{\partial \Phi}{d x^i}\right )_{\rm A}  \simeq 
- \left ( \frac{\partial^2 \Phi}{ \partial x^i\,\partial x^j} 
\right )\,\xi^j \equiv {\cal E}_j^i\,\xi^j\,,
\label{eq:27}
\eeq
where the second equality is obtained by Taylor expanding around 
the position of particle A. The tensor ${\cal E}_j^i$ is called 
the {\it Newtonian tidal-gravity} tensor~\cite{KTcaltech}, it measures the 
inhomogeneities of Newtonian gravity. It is the tensor responsible 
of the Moon's tides on the Earth's ocean.  

We now generalize the above Newtonian discussion to Einstein 
theory. In general relativity, nonspinning test particles move 
along geodesics
\beq
\frac{d^2 x^\mu}{d \tau^2} + \Gamma_{\rho \sigma}^\mu(x)\,
\frac{d x^\rho}{d \tau}\,\frac{d x^\sigma}{d \tau} = 0\,,
\label{eq:28}
\eeq
where $\Gamma_{\rho \sigma}^\mu(x)$ is given by Eq.~(\ref{eq:14}). 
Let us consider two nearby geodesics, labeled A and B, parametrized by 
$x^\mu(\tau)$ and $x^\mu(\tau) + \xi^\mu(\tau)$, with $|\xi^\mu|$ 
smaller than the typical scale on which the gravitational field 
varies. By expanding the geodesic equation of particle B around the 
position of particle A, and subtracting it from the geodesic 
equation of particle A, we get 
\beq
\boldsymbol{\nabla}_u \,\boldsymbol{\nabla}_u \xi^\mu = - R^{\mu}_{\;\;\nu \rho \sigma}\,\xi^\rho \,
\frac{d x^\nu}{d \tau}\,\frac{d x^\sigma}{d \tau}\,,
\label{eq:29}
\eeq
$u^\beta = d x^\beta/d\tau$ being the four-velocity and 
where we introduce the covariant derivative along the curve $x^\mu(\tau)$
\beq
\boldsymbol{\nabla}_u \xi^\mu = \frac{d \xi^\mu}{d \tau} + \Gamma^\mu_{\rho \sigma}\,\xi^\rho\,
\frac{d x^\sigma}{d \tau}\,.
\label{eq:30}
\eeq
Thus, two nearby time-like geodesics experience a tidal gravitational 
{\it force} proportional to the Riemann tensor. 

\subsection{Description in the transverse-traceless gauge}

In this section we describe the interaction of a GW with a point  
particle in the TT gauge. Let us consider a test particle A 
at rest at time $\tau =0$. Using the geodesic equation, we have 
\beq
\frac{d^2 x^i}{d \tau^2}_{|_\tau =0} = - \left ( \Gamma^i_{\rho \sigma}\, 
\frac{d x^\rho}{d \tau}\,\frac{d x^\sigma}{d \tau}\right )_{|_{\tau = 0}} 
= - \left ( \Gamma^i_{00}\,\frac{d x^0}{d \tau}\,\frac{d x^0}{d \tau} 
\right )_{|_{\tau = 0}} \,.
\label{eq:31}
\eeq
Because the particle is initially at rest $(d x^\mu/d \tau)_{\tau =0} = 
(c,0)$ and 
\beq
\Gamma^i_{00} = \frac{1}{2}\,\eta^{i j}\,(\partial_0 h_{0 j} + 
\partial_0 h_{j 0} - \partial_j h_{00} )\,.
\label{eq:32}
\eeq
In the TT gauge $h_{00}=0$ and $h_{0j} =0$, so $(\Gamma^i_{00})_{\tau 
=0}=0$. Thus, we conclude that in the TT gauge, if at time $\tau =0$, 
$d x^i/d \tau = 0$, also $d^2 x^i/d \tau^2 = 0$ and 
a particle at rest before the GW arrives, remains at rest. 
The coordinates in the TT gauge stretch themselves when the 
GW arrives so that the coordinate position of the point 
particles, initially at rest, does not vary. What varies is the proper 
distance between the two particles and physical effects are 
monitored by proper distances. 

For a wave propagating along the $z$-axis, the metric is [see Eq.~(\ref{eq:22})] 
\bea
ds^2 &=& -c^2\,dt^2 + dz^2 + dy^2\,\left [1 - h_+\,\cos \omega\,\left (t - \frac{z}{c} 
\right ) \right ]
+ \nonumber \\
&& dx^2\,\left [1 + h_+\,\cos \omega\,\left (t - \frac{z}{c}\right ) \right ] + 2 dx\,dy\,h_\times\,
\cos \omega\,\left (t - \frac{z}{c} \right )\,.
\eea
If particles A and B set down initially along the $x$-axis, we have 
\beq
s \simeq L\, \left (1 + \frac{h_+}{2}\,\cos \omega t \right )\,,
\label{eq:33}
\eeq
where $L$ is the initial, unperturbed distance between particles A and B. 

\subsection{Description in the free-falling frame}

It is always possible to perform a change of coordinates such that at a given 
space-time point ${\cal Q}$, we can set $\Gamma_{\rho \sigma}^\mu ({\cal Q}) = 0$ 
and  $(d^2 x^\mu/d \tau^2)_{\cal Q} = 0$. In this frame, at one moment 
in space and one moment in time, the point particle is {\it free falling} (FF). This 
frame can be explicitly constructed using {\it Riemann normal coordinates}~\cite{MTW}. 
Actually, it is possible to build a frame such that the point particle 
is free-falling all along the geodesics using {\it Fermi normal coordinates}~\cite{MTW}. 

Let us introduce a FF frame  attached to particle A with spatial 
origin at $x^j=0$ and coordinate time equal to proper time 
$x^0 = \tau$. By definition of a FF frame, the metric reduces to Minkowski 
metric at the origin and all its derivatives vanish at the 
origin, that is  
\beq
ds^2 = - c^2\,d t^2 + d \mathbf{x}^2 + {\cal O}\left ( \frac{|\mathbf{x}|^2}{{\cal R}^2} \right )\,,
\label{eq:34}
\eeq
where ${\cal R}$ is the curvature radius ${\cal R}^{-2} = |R_{\mu \nu \rho \sigma}|$. 
Doing explicitly the calculation at second order in $x$, one finds~\cite{NZ}
\bea
ds^2 &=& -c^2\,dt^2 \, \left [ 1 + R_{i0j0}\,x^i\,x^j \right ] - 2 c\,dt \,d x^i \, \left (
\frac{2}{3} R_{0jik}\,x^j\,x^k \right ) +  \nonumber \\
&&  d x^i\, dx^j \,\left [ \delta_{ij} - \frac{1}{3} 
R_{ ijkl}\,x^k\,x^l \right ]\,.
\label{eq:35}
\eea
For GW experiments located on the Earth, the interferometer is not in free 
fall with respect to the Earth gravity. The detector is subjected to an 
acceleration $\mathbf{a} = - \mathbf{g}$ with respect to a local inertial 
frame and it rotates with respect to local gyroscopes. Thus, in general 
the effect of GWs on point particles compete with other effects. 
We shall restrict our discussion to the frequency band ($10 \mbox{--} 10^3$ Hz) 
in which the other effects are subdominant and/or static. 

Let us compute the equation of geodesic deviation in the FF frame attached to particle A.
We have
\beq
\boldsymbol{\nabla}_u\,\boldsymbol{\nabla}_u \xi^\alpha = u^\beta\,\boldsymbol{\nabla}_\beta (u^\lambda \,\boldsymbol{\nabla}_\lambda \xi^\alpha) = 
u^\beta\,u^\lambda\,(\partial_{\beta \lambda} \xi^\alpha + \Gamma_{\lambda \sigma, \beta}^\alpha\,\xi^\sigma )\,,
\label{eq:36}
\eeq
where in the last equality we use $\Gamma_{\lambda \sigma}^\alpha = 0$. 
Since we assume that the particles are initially at rest, $u^\beta = \delta_0^\beta$.  
Using $\xi^0 =0$ and the fact that $\Gamma^j_{0k,0}$ can be neglected when computed 
at position A, we have
\beq
\boldsymbol{\nabla}_u\,\boldsymbol{\nabla}_u \xi^j = \frac{d^2 \xi^j}{d \tau^2} 
\quad \Rightarrow \quad \frac{d^2 \xi^j}{d \tau^2} = - R^j_{\;0k0}\,\xi^k\,.
\label{eq:37}
\eeq
To complete the calculation we need to evaluate the Riemann 
tensor $R^j_{\;0k0}$. As discussed in Sec.~\ref{sec2.2}, 
in linearized theory the Riemann tensor is invariant 
under change of coordinates, so we can compute it in the TT gauge. 
Using Eq.~(\ref{eq:15}), we obtain 
\beq
R_{j0k0}^{\rm TT} = - \frac{1}{2c^2}\,\ddot{h}_{jk}^{\rm TT}\,.
\label{eq:38}
\eeq
Thus, 
\beq
\frac{d^2 \xi^j}{d t^2} = \frac{1}{2}\ddot{h}_{jk}^{\rm TT}\,\xi^k\,.
\label{eq:39}
\eeq
In conclusion, in the FF frame the effect of a GW on a point particle 
of mass $m$ can be described in terms of a {\it Newtonian force} 
$F_i = (m/2)\,\ddot{h}^{\rm TT}_{ij}\,\xi^j$. Note that in the FF frame, 
coordinate distances and proper distances coincide, and we recover 
immediately Eq.~(\ref{eq:33}). 

The description in the FF frame is useful and simple as long as we 
can write the metric as $g_{\mu \nu} = \eta_{\mu \nu} + {\cal O}(x^2/{\cal R}^2)$, 
i.e. as long as we can disregard the corrections $x^2/{\cal R}^2$. 
Since ${\cal R}^{-2} = |R_{i0j0}| \sim \ddot{h} \sim h/\lambda_{\rm GW}^2$, 
we have $x^2/{\cal R}^2 \simeq L^2\,h/\lambda_{\rm GW}^2$, and comparing 
it with $\delta L/L \sim h$, we find $L^2/\lambda_{\rm GW}^2 \ll 1$. This 
condition is satisfied by ground-based detectors because $L \sim 4$ km 
and $\lambda_{\rm GW} \sim 3000$ km, but not by space-based detectors 
which have $L \sim 5 \times 10^6$ km  and will observe GWs with 
wavelength shorter than $L$. [For a recent thorough analysis and a proof of the 
equivalence between the TT and FF description, see, e.g., Ref.~\cite{MR}.]  
\begin{figure}
\begin{center}
\begin{tabular}{cc}
\includegraphics[width=0.45\textwidth]{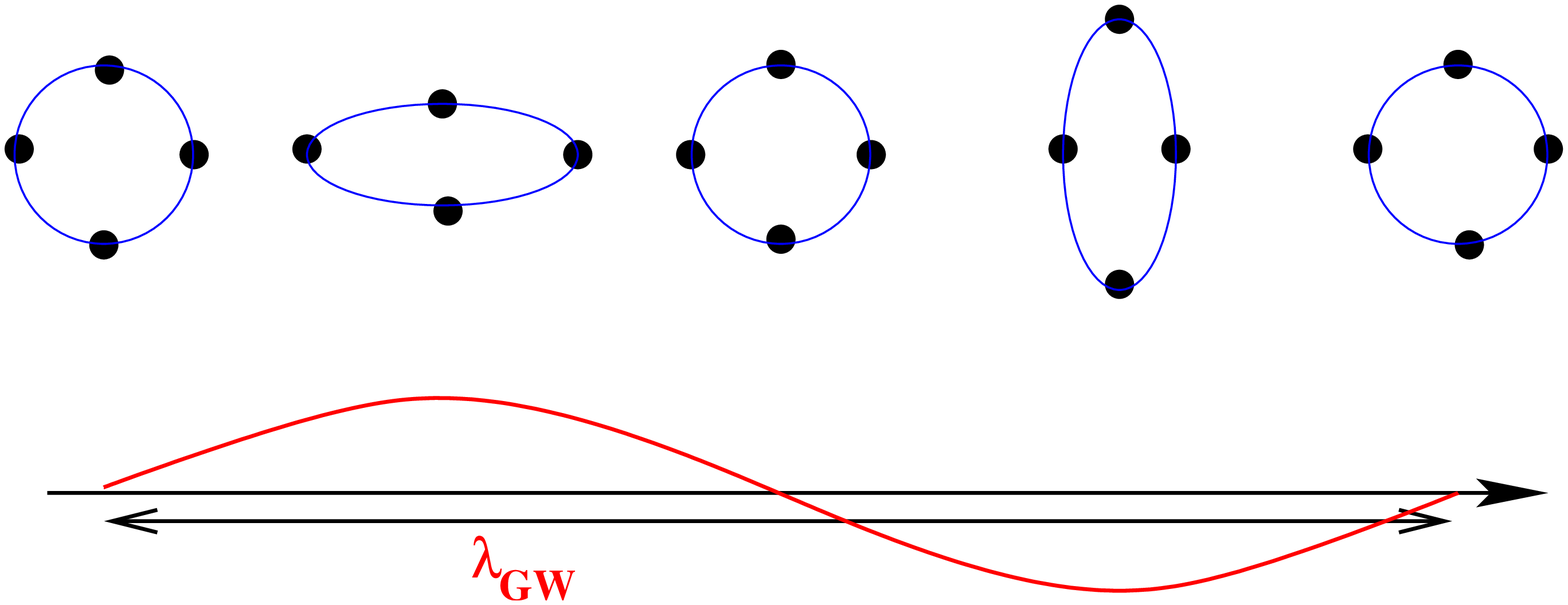} &
\includegraphics[width=0.45\textwidth]{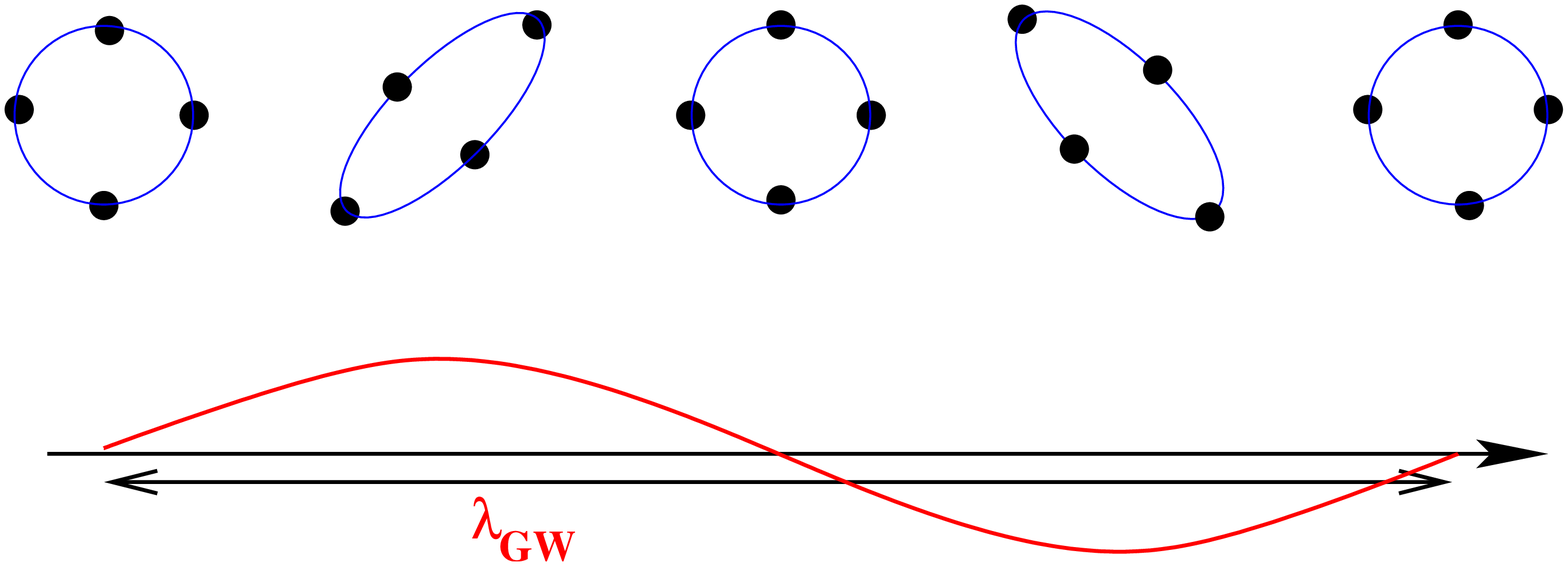}\\
\end{tabular}
\caption{We show how point particles along a ring move as a result of 
the interaction  with a GW propagating in the direction perpendicular to the 
plane of the ring. The left panel refers to a wave 
with $+$ polarization, the right panel with $\times$ polarization. 
\label{fig:2}}
\end{center}
\end{figure}
\subsection{Key ideas underlying gravitational-wave detectors}

\begin{figure}
\begin{center}
\begin{tabular}{cc}
\includegraphics[width=0.45\textwidth]{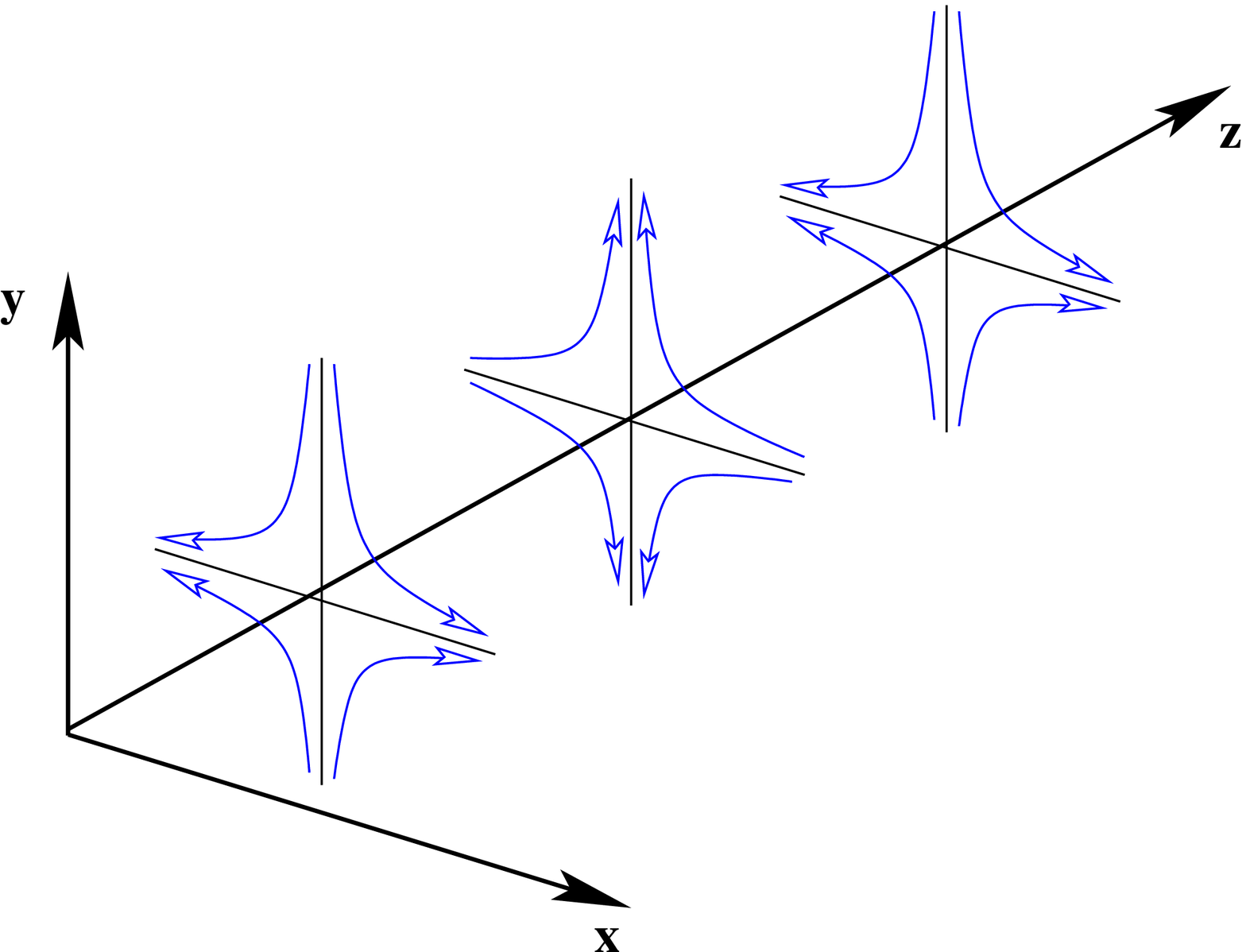} &
\includegraphics[width=0.45\textwidth]{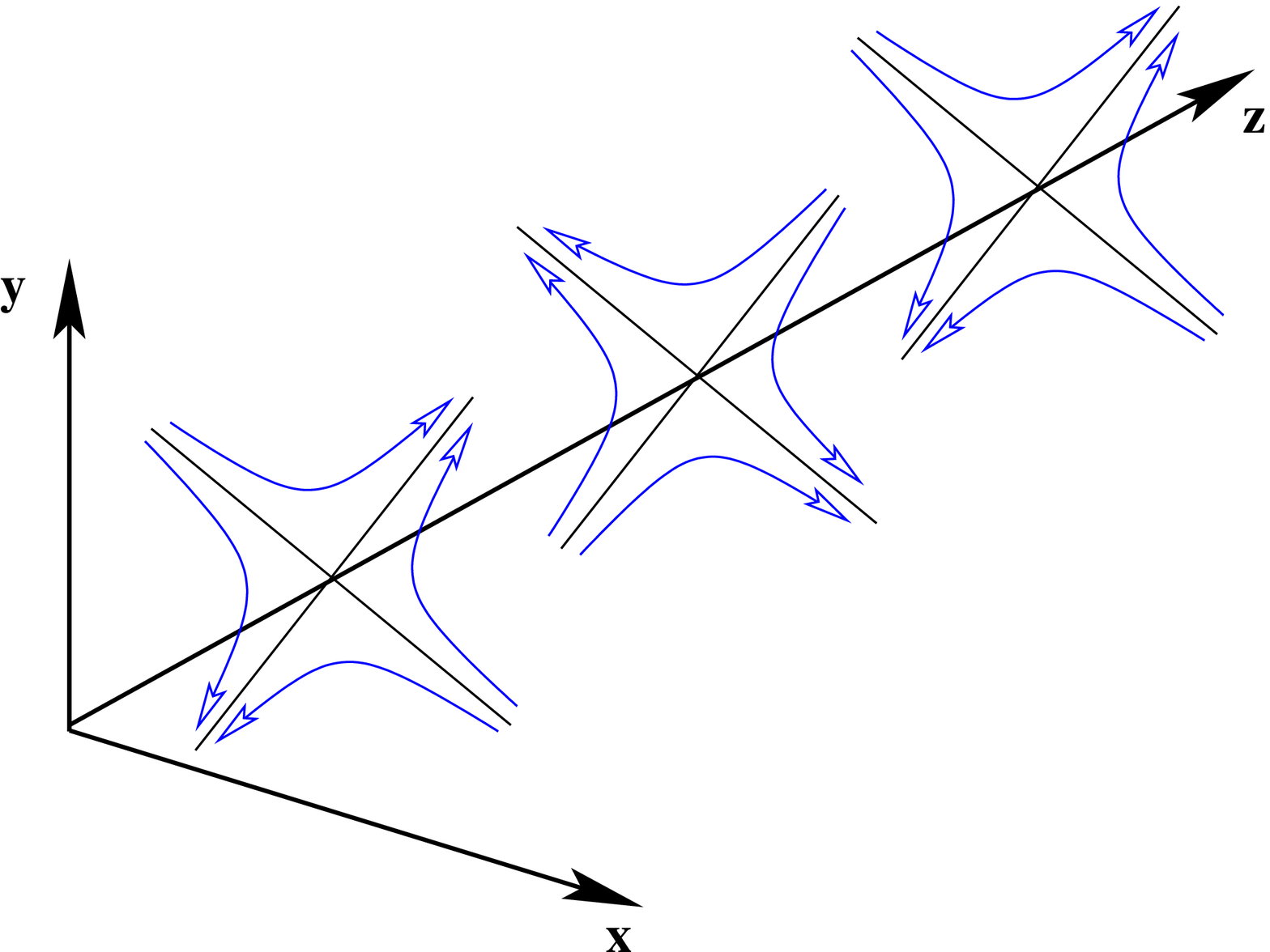} \\
\end{tabular}
\caption{Lines of force associated to the $+$ (left panel) and $\times$ (right panel) polarizations.\label{fig:3}}
\end{center}
\end{figure}

To illustrate the effect of GWs on FF particles, 
we consider a ring of point particles initially at rest with respect to a 
FF frame attached to the center of the ring, as shown in Fig.~\ref{fig:2}. 
We determine the motion of the particles 
considering the $+$ and $\times$ polarizations separately.
If only the $+$ polarization is present, we have 
\beq
h_{i j}^{\rm TT} = h_+ \left ( \begin{array}{cc}
1 & 0 \\
0 & -1 
\end{array} \right ) \sin \omega t\,, \quad 
\xi_i = [x_0 + \delta x(t),y_0 + \delta y(t)]\,,
\label{eq:40}
\eeq
where $x_0$ and $y_0$ are the unperturbed position at time $t=0$. Thus 
\beq
\delta x(t) = \frac{h_+}{2}\,x_0\,\sin \omega t \quad \quad 
\delta y(t) = - \frac{h_+}{2}\,y_0\,\sin \omega t \,.
\label{eq:41}
\eeq
If only the $x$ polarization is present, a straightforward calculation 
gives
\beq
\delta x(t) = \frac{h_\times}{2}\,y_0\,\sin \omega t \quad \quad 
\delta y(t) = \frac{h_\times}{2}\,x_0\,\sin \omega t\,.
\label{eq:42}
\eeq
The $+$ and $\times$ polarizations differ by a rotation of $45^o$. 
In Fig.~\ref{fig:3} we show the lines of force associated to the 
$+$ and $\times$ polarizations. 

The simplest GW detector we can imagine is a body of mass $m$ at a 
distance $L$ from a fiducial laboratory point, connected to it 
by a spring of resonant frequency $\Omega$ and quality 
factor $Q$. Einstein equation of geodesic deviation predicts that 
the infinitesimal displacement $\Delta L$ of the mass  along the 
line of separation from its equilibrium position satisfies the 
equation~\cite{KT87} (valid for wavelengths $\gg L$ and 
in the FF frame of the observer at the fiducial 
laboratory point) 
\beq
\ddot{\Delta L}(t) + 2\frac{\Omega}{Q}\,\dot{\Delta L}(t) + \Omega^2\,\Delta L(t) = 
\frac{L}{2}\,\left [F_+\,\ddot{h}_+(t) + F_\times\,\ddot{h}_\times(t) \right ]\,,
\label{sd}
\eeq
where $F_{+,\times}$ are coefficients of order unity which depend on the 
direction of the source [see Eqs.~(\ref{F+}), (\ref{Fx}) below] and the 
GW polarization angle. 

Laser-interferometer GW detectors 
are composed of two perpendicular km-scale arm cavities with two 
test-mass mirrors hung by wires at the end of each cavity.
The tiny displacements $\Delta L$ of the mirrors induced by a passing GW 
are monitored with very high accuracy by measuring the relative optical phase 
between the light paths in each interferometer arm.
The mirrors are pendula with quality factor $Q$ quite high and 
resonant frequency $\Omega$ much lower ($\sim 1$ Hz) than the typical 
GW frequency ($\sim 100$ Hz). In this case Eq.~(\ref{sd}), written 
in Fourier domain, reduces to $\Delta L/L \sim h$. 
The typical amplitude, at $100$ Hz, of GWs emitted by binary systems 
in the VIRGO cluster of galaxies ($\sim 20$ Mpc distant), 
which is the largest distance the first-generation  
ground-based interferometers can probe, is $\sim 10^{-21}$. 
This means $\Delta L \sim 10^{-18}$ m, a very tiny number. 
It may appear rather discouraging, especially if we think to monitor 
the test-mass motion with light of wavelength nearly $10^{12}$ times 
larger. However, this precision is currently be demonstrated 
experimentally. 

\begin{figure}
\begin{center}
\begin{tabular}{c}
\includegraphics[width=0.8\textwidth]{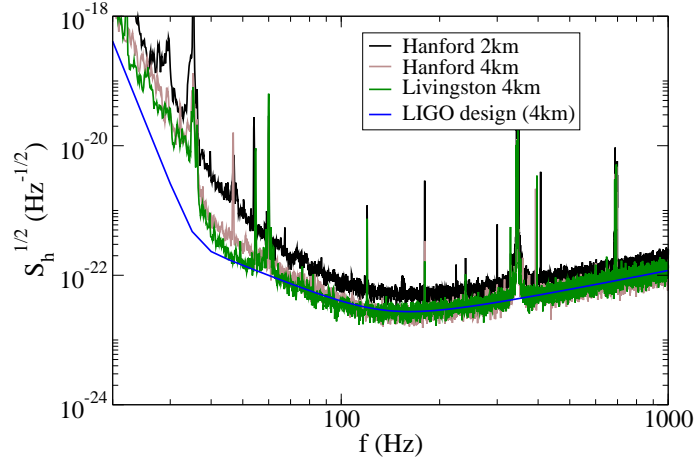}
\end{tabular}
\caption{We plot the square root of the noise spectral density 
versus frequency for the three LIGO detectors together with the 
LIGO noise curve at designed sensitivity . The noise 
curves refer to June 2006, during the fifth scientific run. \label{fig:4}}
\end{center}
\end{figure}

The electromagnetic signal leaking out the interferometer's dark-port 
contains the GW signal but also noise --- 
for example the thermal noise from the suspension system and 
the mirror itself, can shake the mirror mimicking the effect 
of a GW. The root-mean-square of the noise is generally  expressed in terms 
of the noise power per unit frequency $S_h$ through the relation 
$h \sim \sqrt{S_h(f)\,\Delta f} \sim \Delta L/L$, $\Delta L$ being the 
mirror displacement induced by noise and $\Delta f$ the frequency bandwidth. 
In Fig.~\ref{fig:4} we plot the noise curves of LIGOs (June 2006). 
The interferometers are currently operating at design sensitivity 
for almost the entire frequency band. 

\section{Effective stress-energy tensor of gravitational waves}
\label{sec4}

Until now we have defined the GWs as fluctuations of a flat spacetime. 
Here, we want to be more general and consider GWs as perturbations of a generic 
background $\overline{g}_{\mu \nu}$, that is 
\beq
g_{\mu \nu} = \overline{g}_{\mu \nu} + h_{\mu \nu}\,, \quad \quad |h_{\mu \nu}| \ll 1\,.
\label{eq:43}
\eeq
We need a criterion to define what is the background and what is the perturbation. 
Heenceforth, we follow closely the derivation in Ref.~\cite{M07}. 
In general there are two cases: 
\begin{enumerate}
\item $\overline{g}_{\mu \nu}$ has typical scale $L_{\rm B}$ and 
$h_{\mu \nu}$ has typical wavelength $\lambda$ with $\lambda \ll L_{\rm B}$, 
i.e. $h_{\mu \nu}$ is a small {\it ripple} on a smooth background; 
\item $\overline{g}_{\mu \nu}$ has frequencies only up to $f_B$ and  
$h_{\mu \nu}$ is different from zero around $f$ with 
$f \gg f_{\rm B}$, i.e. the background is static.
\end{enumerate}
Let us expand $R_{\mu \nu}$ through ${\cal O}(h^2)$. Note that 
we have now two small parameters $h$ and $\lambda/L_{\rm B}$ (or $f_{\rm B}/f$), 
we have 
\beq
R_{\mu \nu} = \underbrace{\overline{R}_{\mu \nu}}_{\rm low\,freq} \; + \;
\underbrace{{R}_{\mu \nu}^{(1)}}_{\rm high\,freq} \; + \; 
\underbrace{{R}_{\mu \nu}^{(2)}}_{\rm low\, and\, high\, freq.} + \cdots \,.
\label{eq:44}
\eeq
Quantities having a bar are computed using the background metric 
$\overline{g}_{\mu \nu}$; they contain only low-frequency 
modes. The superscript $(1)$ [$(2)$] in Eq.~(\ref{eq:44}) refers to quantities 
computed at linear (quadratic) order in $h$. Using the Einstein equations we get 
\beq
\overline{R}_{\mu \nu} = - [R^{(2)}_{\mu \nu} ]^{\rm low \, freq} + 
\frac{8 \pi G}{c^4}\,\left [ T_{\mu \nu} - \frac{1}{2}g_{\mu \nu}\,T \right ]^{\rm low \, freq}\,.
\label{eq:45}
\eeq
We introduce a scale $\ell$ such that $\lambda \ll \ell \ll L_{\rm B}$, 
and average over a spatial volume $\ell^3$~\cite{M07,average}. We denote the average 
as $\langle \rangle$. 
Short-wave modes average to zero, whereas 
modes with wavelength $L_B$ are constant. We can rewrite Eq.~(\ref{eq:45}) as
\bea
\overline{R}_{\mu \nu} &=& - \langle R^{(2)}_{\mu \nu} \rangle + \frac{8 \pi G}{c^4} \,
\langle T_{\mu \nu} - \frac{1}{2}\,g_{\mu \nu}\,T \rangle \nonumber \\ 
& \equiv & - \langle R^{(2)}_{\mu \nu} \rangle + \frac{8 \pi G}{c^4} \,
\left (\overline{T}_{\mu \nu} - \frac{1}{2}\,\overline{g}_{\mu \nu}\,\overline{T} \right )\,.
\label{eq:46}
\eea
Defining the {\it effective} stress-energy tensor of GWs 
\beq
t_{\mu \nu} = -\frac{c^4}{8 \pi G}\,\langle R^{(2)}_{\mu \nu} - \frac{1}{2}\,\overline{g}_{\mu \nu}\,R^{(2)} \rangle\,,
\label{eq:47}
\eeq
we have 
\beq
\overline{R}_{\mu \nu} - \frac{1}{2}\,\overline{g}_{\mu \nu}\,\overline{R} = 
\frac{8 \pi G}{c^4}\,(\overline{T}_{\mu \nu} + t_{\mu \nu}).
\label{eq:48}
\eeq
An explicit calculation carried on far from the source gives
\beq
t_{\mu \nu} = \frac{c^4}{32 \pi G}\,\langle \partial_\mu h_{\alpha \beta}\,\partial_\nu h^{\alpha 
\beta}\rangle\,.
\label{eq:49}
\eeq
For a plane wave, using the TT gauge
\beq
t_{00} = \frac{c^2}{32 \pi G}\,\langle \dot{h}^{\rm TT}_{ij}\,\dot{h}^{\rm TT}_{ij} \rangle = 
\frac{c^2}{16 \pi G}\,\langle \dot{h}_+^2 + \dot{h}_\times^2 \rangle\,,
\label{eq:50}
\eeq
and the GW energy flux per unit area is 
\beq
\frac{d E}{dt\,dA} = \frac{c^3}{16 \pi G}\,\langle \dot{h}_+^2 + \dot{h}_\times^2 \rangle\,.
\label{eq:51}
\eeq
For a supernovae ${d E}/(dt\,dA) \sim c^3 f^2 h^2/(16 \pi G) \sim 400 {\rm erg}/({\rm cm}^2 {\rm sec})$,  
where we set $h=10^{-21}$ and $f = 1$ kHz. The GW burst has a duration of a few msec. 
It is telling to compare it with the neutrino energy flux $\sim 10^5 
{\rm erg}/({\rm cm}^2 {\rm sec})$ and the photon energy flux (optical radiation) 
$\sim 10^{-4} {\rm erg}/({\rm cm}^2 {\rm sec})$ from a supernovae. 
Neutrinos and optical radiation are emitted during a few seconds and one week, respectively.

\section{Generation of gravitational waves}
\label{sec5}

\subsection{Sources in slow motion, weak-field and negligible self-gravity}
\label{sec5.1}

In this section we evaluate the leading order contribution to the metric 
perturbations under the assumption that the internal motions of the 
source are slow compared to the speed of light. We also assume that the 
source's self-gravity is negligible. Henceforth, we shall discuss how to 
extend those results to sources with non-negligible self-gravity. 
We start from 
\beq
\Box \overline{h}_{\mu \nu} = 
- \frac{16 \pi G}{c^4}\,T_{\mu \nu}\,, \quad \partial_\mu 
\overline{h}^{\mu \nu} = 0\,, \quad \partial_\mu T^{\mu \nu} = 0\,,
\label{eq:52}
\eeq
and introduce retarded Green functions
\beq
G(x - x') = - \frac{1}{4 \pi}\,\frac{1}{|\mathbf{x} - \mathbf{x}'|}\,
\delta \left ( t - \frac{|\mathbf{x} - \mathbf{x}'|}{c} - t' \right )\,, 
\label{eq:53}
\eeq
which satisfy $ \Box_x G(x - x') = \delta^{(4)}(x - x')$. 
The solution of Eq.~(\ref{eq:52}) can be written as 
\beq 
\overline{h}_{\mu \nu}(x)  = - \frac{16 \pi G}{c^4}\,\int d^4 x'\,G(x - x')\,T_{\mu \nu}(x')\,.
\label{eq:54}
\eeq
Outside the source, using the TT gauge, we have  
\beq 
\overline{h}_{i j}^{\rm TT}(t,\mathbf{x})  = \Lambda_{ij,kl}(\mathbf{n})\,
\frac{4 G}{c^4}\,\int d^3 x'\,\frac{1}{|\mathbf{x}-\mathbf{x}'|}\,
T_{k l}\left (t - \frac{|\mathbf{x} - \mathbf{x}'|}{c};\mathbf{x}' \right )\,.
\label{eq:55}
\eeq
Denoting by $d$ the typical size of the source, assuming to be far 
from the source, i.e. $r \gg d$, we can write 
$|\mathbf{x} - \mathbf{x}'| = r - \mathbf{x}'\cdot \mathbf{n} + {\cal O}(d^2/r)$, 
and Eq.~(\ref{eq:54}) becomes
\beq
\overline{h}_{i j}^{\rm TT}(t,\mathbf{x})  \simeq \frac{1}{r}\, 
\frac{4 G}{c^4}\,\Lambda_{ij,kl}(\mathbf{n})\,
\int_{|\mathbf{x}'|<d}  d^3 x' \, T_{k l}\left (t - \frac{r}{c} + \frac{\mathbf{x}'\cdot \mathbf{n}}{c};\mathbf{x}' \right )\,.
\label{eq:56}
\eeq
We can simplify the above equations if we assume that typical velocities inside 
the sources are much smaller than the speed of light $c$. If $\omega$ is 
the typical frequency associated to the source motion, typical 
source velocities are $v \sim \omega\,d$. As we shall see in the following, 
the GW signal is determined by the leading 
multipole moments, thus $\omega_{\rm GW} \sim \omega \sim v/d$ 
and $\lambda_{\rm GW} \sim (c/v)\,d$. If $v/c \ll 1$, we have 
$\lambda_{\rm GW} \gg d$. 

Applying a Fourier decomposition, we can write 
\beq
T_{k l} \left ( t - \frac{r}{c} + \frac{\mathbf{x}' \cdot n}{c}; \mathbf{x}' \right ) = 
\int \frac{d^4 k}{(2 \pi)^4}\, \tilde{T}_{kl}(\omega, \mathbf{k}) \times 
e^{- i \omega\,\left (t - \frac{r}{c} + \frac{\mathbf{x}'\cdot \mathbf{n}}{c} \right ) + i \mathbf{k}\cdot \mathbf{x}'}\,,
\label{eq:57}
\eeq
using $\omega \mathbf{x}'\cdot \mathbf{n} \sim \omega\,d/c \ll 1$, expanding the 
exponential and Taylor-expanding $T_{kl}$ we get
\bea
h_{i j}^{\rm TT}(t, \mathbf{x}) &\simeq& \frac{1}{r}\,\frac{4 G}{c^4}\,\Lambda_{ij, kl}(\mathbf{n})\,
\left [ \int d^3x\, T^{k l}(t, \mathbf{x}) + \right . \nonumber \\
&& \left . \frac{1}{c}\,n_m\, \frac{d}{d t} \int d^3 x\, T^{ kl}(t, \mathbf{x})\,x^m + 
\right . \nonumber \\
&& \left . \frac{1}{2 c^2}\,n_m\,n_p\,\frac{d^2}{d t^2} \int d^3 x\, T^{kl}(t,\mathbf{x})\,x^m\,x^p + 
\cdots \right ]_{|_{t - r/c}}\,.
\label{eq:58}
\eea
The above expression is valid in linearized gravity and for negligible 
self-gravity sources, i.e. for sources whose dynamics is not determined by 
gravitational forces. We notice that in Eq.~(\ref{eq:58}) the higher multipoles are suppressed 
by a factor $v/c$. To make Eq.~(\ref{eq:58}) more transparent,  
we can express the momenta $T^{ij}$ in terms of the momenta of 
$T^{00}$ and $T^{0i}$. Let us first introduce the momenta of the mass density 
\bea
M &=& \frac{1}{c^2} \int d^3 x\, T^{00}(t, \mathbf{x})\,, \\
M^i &=& \frac{1}{c^2} \int d^3 x\, T^{00}(t, \mathbf{x})\,x^i\,, \\
M^{i j} &=& \frac{1}{c^2} \int d^3 x\, T^{00}(t, \mathbf{x})\,x^i\,x^j\,, \\
\label{eq:59}
\eea
and impose the conservation law $\partial_\mu T^{\mu \nu} = 0$ valid in linearized gravity. 
Setting $\nu=0$, we have $\partial_0 T^{00} + \partial_i T^{i0} = 0$, integrating this equation 
in a volume containing the source, we obtain the conservation of the mass $\dot{M}=0$. 
Similarly, one can prove the conservation of the momentum $\ddot{M}^i = 0$ 
Moreover, we have 
\bea
c\, \dot{M}^{ij} &=& \int_V d^3 x \, x^i \,x^j \partial_0 T^{00} = 
- \int_V d^3 x \, x^i \, x^j \partial_k T^{0k}  \nonumber \\
&=& \int_V d^3x\, (x^j\,T^{0i} + x^i\,T^{0j})\,,
\label{eq:60}
\eea
where the second line is obtained after integrating by parts. Finally, 
\beq 
\ddot{M}^{ij} = 2 \int_V d^3 x \, T^{ij}\,.
\label{eq:61}
\eeq
Thus, the leading term in Eq.~(\ref{eq:58}), can be rewritten as 
\beq
h^{\rm TT}_{ij}(t,\mathbf{x}) = \frac{1}{r}\,\frac{2 G}{c^4}\,\Lambda_{ij,kl}(\mathbf{n})\,
\ddot{M}^{kl}\left ( t - \frac{r}{c}\right )\,, 
\label{eq:62}
\eeq
where $M^{kl}$ is given by Eq.~(\ref{eq:59}). The quantity $T^{00}/c^2$ in Eq.~(\ref{eq:59}) 
is a mass density. Besides the rest-mass contribution, it can contain terms due 
to the kinetic energy and the potential energy. For sources having strong 
gravitational field, as NSs and BHs, the mass density can depend also on the 
binding energy. Only for weak fields and small velocities, which is 
the assumption so far made, $T^{00}/c^2$  reduces to the rest-mass 
density $\rho$. 

Henceforth, we shall discuss 
some applications to binary systems and pulsars, so it is convenient to compute 
explicitly $h_+$ and $h_\times$. Assuming that the GW propagates along the direction $\mathbf{n} = 
(\cos \phi\, \sin \theta, \sin \phi\,\sin \theta, \cos \theta)$, a straight 
calculations gives:
\bea
h_+ &=& \frac{G}{r\,c^4}\, \left \{ 
\ddot{M}_{11}\,(\sin^2 \phi - \cos^2 \theta\,\cos^2 \phi) 
+ \ddot{M}_{22}\,(\cos^2 \phi - \cos^2 \theta\,\sin^2 \phi) 
\right . \nonumber \\
&& \left. - \ddot{M}_{33}\,\sin^2 \theta - \ddot{M}_{12}\, \sin 2 \phi \, 
(1+ \cos^2 \theta) + \ddot{M}_{13}\, \cos \phi\, \sin 2 \theta + \right. \nonumber \\
&& \left. \ddot{M}_{23}\,\sin 2 \theta\, \sin \phi \right \}\,,
\label{eq:63}
\eea
\bea
h_\times &=& \frac{2 G}{r\,c^4}\, \left \{ 
\frac{1}{2}\,(\ddot{M}_{11} - \ddot{M}_{22})\,\cos \theta\,\sin 2\phi  - 
\ddot{M}_{12}\,\cos \theta\, \cos 2\phi \right. \nonumber \\
&& \left. - \ddot{M}_{13}\,\sin \theta\, 
\sin \phi + \ddot{M}_{23} \, \cos \phi \sin \theta \right \}\,.
\label{eq:64}
\eea

\subsection{Sources in slow motion and weak-field, but non-negligible self-gravity}
\label{sec5.2}

As already stated above, the derivation in linearized gravity of the quadrupole 
formula (\ref{eq:62}) cannot be applied to systems like binary stars whose dynamics 
is dominated by gravitational forces. In fact, because of the conservation-law 
valid in linearized gravity $\partial_\mu T_{\mu \nu} = 0$, the two bodies 
move along geodesics in Minkowski spacetime. The extension to the case 
in which self-gravity is non-negligible can be done as follows~\cite{Quad}. 

In full general relativity one can define the field ${H}_{\mu \nu}$ 
such that 
\beq
\sqrt{-g}\,g^{\mu \nu} = \eta^{\mu \nu} - {H}^{\mu \nu}\,,
\label{eq:65}
\eeq
where in the weak-field limit ${H}^{\mu \nu}$ coincides with the 
reverse-trace tensor that we introduced  above. Imposing the harmonic 
gauge $\partial_\mu {H}^{\mu \nu}=0$, one derives 
\beq
\Box {H}_{\mu \nu} = - \frac{16 \pi G}{c^4}\,
\left [ (-g)\,T_{\mu \nu} + \tau_{\mu \nu} \right ]\,,
\label{eq:66}
\eeq
where $\tau_{\mu \nu}$ is the pseudotensor depending on $H_{\mu \nu}$ that 
can be read explicitly from Refs.~\cite{LL,PNreview}. 
The conservation law reads in this case
\beq
\partial_\mu \left [(-g)\,T^{\mu \nu} + \tau^{\mu \nu} \right ]=0\,.
\label{eq:69}
\eeq
We can redo the derivation in linearized gravity (see Sec.~\ref{sec2.2}), but  
replace $T_{\mu \nu} \rightarrow (-g)\,T_{\mu \nu} + \tau_{\mu \nu}$, 
obtaining for the leading term in Eq.~(\ref{eq:62}) 
$\int T_{00}\,x^i\,x^j\,d^3 x \rightarrow \\ \int (T_{00}+ 
\tau_{00})\,x^i\,x^j\,d^3 x $. For sources characterized 
by weak gravity $\tau_{00}$ is negligible with respect to 
$T_{00}$. Even though at the end one obtains the same formula, 
it is crucial to take into account the second order corrections in 
$h_{\mu \nu}$, i.e. the field $\tau_{\mu \nu}$, in the conservation 
law. Otherwise the sources would be obliged to move along 
geodesics in Minkowski spacetime, instead of moving in a 
bounded orbit.

\subsection{Radiated energy, angular momentum  and linear momentum}
\label{sec5.3}

Using the results of Sec.~\ref{sec4}, we can compute the power radiated 
at leading order
\beq
\frac{d P}{d \Omega } = \frac{r^2\,c^3}{32 \pi\,G} 
\langle \dot{h}^{\rm TT}_{i j}\,\dot{h}^{\rm TT}_{i j} \rangle 
= \frac{G}{8 \pi\,c^5}\,\Lambda_{kl,mp}(\mathbf{n})\,\langle \dddot{Q}_{kl}\,
\dddot{Q}_{mp} \rangle\,,
\label{eq:70}
\eeq
where we introduce the traceless quadrupole tensor
\beq
Q_{ij} = M_{ij} - \frac{1}{3}\,\delta_{ij}\,M_{kk}\,. 
\eeq
Using the following relations
\beq
\int \frac{d \Omega}{4 \pi}\,n_i\,n_j = \frac{1}{3}\,\delta_{ij}\,,
\label{eq:71}
\eeq
\beq
\int \frac{d \Omega}{4 \pi}\,n_i\,n_j\,n_k\,n_l = 
\frac{1}{15}\,(\delta_{ij}\,\delta_{kl} + 
\delta_{ik}\,\delta_{jl} + \delta_{il}\,\delta_{jk})\,,
\label{eq:72}
\eeq
we derive for the total power radiated 
\beq
P = \frac{G}{5 c^5}\, \langle \dddot{Q}_{ij}\, \dddot{Q}_{ij} \rangle\,. 
\label{eq:73}
\eeq
In the literature Eq.~(\ref{eq:73}) is generally denoted as the 
{\it quadrupole formula}. 

GWs not only carry away from the source the energy, but also angular momentum and linear momentum. 
At leading-order the angular-momentum radiated is~\cite{M07} 
\beq
\frac{d L^i}{dt } = \frac{2 G}{5 \,c^5}\,\epsilon^{ijk}\, \langle \ddot{Q}_{jl}\,\dddot{Q}_{lk} 
\rangle
\eeq
while the linear momentum radiated is given by~\cite{M07}  
\beq
\frac{d P^i}{d t} = - \frac{G}{8 \pi\,c^5}\, 
\int d \omega\,\dddot{Q}^{\rm TT}_{jk}\,\partial^i \ddot{Q}_{jk}^{\rm TT}\,.
\label{eq:74}
\eeq
Under parity, $\mathbf{x} \rightarrow - \mathbf{x}$, the mass quadrupole 
does not vary, and the integral in Eq. (\ref{eq:74}) is overall odd and vanishes. The 
first nonzero contribution comes at order ${\cal O}(1/c^7)$ from 
the interference between the mass quadrupole and the sum of the 
octupole and current quadrupole. As a consequence of the loss of 
linear momentum through GW emission, the BH formed by the 
coalescence of a BH binary can acquire a kick or recoil 
velocity. The recoil velocity is astrophysically significant. 
If it were too large, the BH can be ejected from the host galaxy with 
important consequences on BH's mass growth through hierarchical mergers.  
Recently, there have been a plethora of numerical~\cite{recoilNR} 
and analytic~\cite{recoilAN} predictions. 

The above discussion on energy, angular momentum and linear-momentum 
can be made more rigorous and applicable to sources 
with non-negligible self gravity thanks to the work of Bondi in the 
50s~\cite{HB}. Let us consider a sphere ${\cal S}$ 
of volume ${\cal V}$ and radius $r$ containing the source. 
Be $r$ much larger than the source dimension and 
the gravitational wavelength (far zone). 
It can be proven~\cite{HB,LL} that $P^{\mu}$, 
defined by~\footnote{Note that 
the integration is done over a constant-time hypersurface.}
\beq
P^\mu = \int \tau^{\mu 0}\,d^3 x\,,
\label{eq:75}
\eeq
is a four vector with respect to Lorentz transformations. 
Using the relation $\partial_\mu \tau^{\mu \nu} = 0$, we can write 
\beq
\frac{d P^\mu}{dt} = \int_{\cal V} \partial_0 \tau^{\mu 0}\,d^3 x
= - \oint_{\cal S} \tau^{\mu i}\,n^i\,dS\,.
\label{eq:76}
\eeq
For $\mu=0$ the above equation gives the conservation of 
the energy
\beq
\frac{d P^0}{d t} = - r^2\,\oint d \Omega\, \tau^{0i}\,n_i\,,
\label{eq:77}
\eeq
for $\mu=j$ Eq.~(\ref{eq:76}) gives the conservation of 
the linear momentum 
\beq
\frac{d P^j}{d t} = - r^2\,\oint d \Omega\, \tau^{ji}\,n_i\,.
\label{eq:78}
\eeq

\section{Application to binary systems}
\label{sec6}

\subsection{Inspiral waveforms at leading Newtonian order}
\label{sec6.1}

Let us consider a binary system with masses $m_1$ and $m_2$, 
total mass $M=m_1+m_2$, reduced mass $\mu = m_1\,m_2/(m_1+m_2)$ and symmetric mass-ratio 
$\nu = \mu/M$. We first assume that the two bodies are rather 
separated and move along a circular orbit. In the center-of-mass frame we can write 
for the relative coordinates
\beq
X(t) = R\,\cos \omega\,t \,, \quad Y(t) = R\,\sin \omega\,t \,, \quad Z(t) =0\,,
\label{eq:79}
\eeq
$R$ being the relative distance between the two bodies. 
The only nonzero components of the tensor $M^{ij} = \mu X^i\,X^j$ are 
\bea
M_{11} &=& \frac{1}{2}\,\mu\,R^2\,(1+\cos 2 \omega t )\,,\\
M_{22} &=& \frac{1}{2}\,\mu\,R^2\,(1-\cos 2 \omega t )\,,\\
M_{12} &=& \frac{1}{2}\,\mu\,R^2\,\sin 2 \omega t \,.
\label{eq:80}
\eea
Taking time-derivatives and plugging the above expressions in Eqs.~(\ref{eq:62}), (\ref{eq:63})  
we obtain
\bea
h_+(t) &=& \frac{1}{r}\,\frac{4G}{c^4}\,\mu\,R^2\,\omega^2\,\frac{(1+\cos^2 \theta)}{2}\,\cos (2 \omega\,t)\,
\,,\\
h_\times(t) &=& \frac{1}{r}\,\frac{4G}{c^4}\,\mu\,R^2\,\omega^2\,\cos \theta\,\sin (2 \omega\,t)\,,
\label{eq:81}
\eea
where we shift the origin of time to get rid of the dependence on $\phi$ and trade 
the retarded time with $t$. 

For $\theta = 0$, i.e. along the direction perpendicular to the orbital plane, $h_+$ 
and $h_\times$ are both different from zero, and $h_\times \pm i\,h_+ \propto \pm i\,e^{-2i\omega t}$, 
thus the wave is circularly polarized. For $\theta = \pi/2$, i.e. along the orbital plane, 
only $h_+$ is different from zero and the wave is linearly polarized. 

The angular distribution of the radiated power is given by Eq.~(\ref{eq:73}). It reads
\bea
\left ( \frac{dP}{d \Omega} \right ) &=& \frac{2 G\,\mu^2\,R^4\,\omega^6}{\pi c^5}\,
{\cal P}(\theta)\,,\\
{\cal P}(\theta) &=& \frac{1}{4}\,(1 + 6\,\cos^2 \theta + \cos^4 \theta)\,.
\label{eq:82}
\eea
The maximum power is emitted along the direction perpendicular to the 
orbital plane, $\theta=0$, where ${\cal P}(\pi/2) = 2$. Since 
in the case of a binary system, there is always a component of the source's motion 
perpendicular to the observation direction, the power radiated does not 
vanish in any direction. Integrating over the total solid angle, we obtain 
the total power radiated, 
\beq
P = \frac{32}{5}\,\frac{G\,\mu^2\,R^4\,\omega^6}{c^5}\,.
\label{eq:83}
\eeq
If we consider the binary system composed of Jupiter and the Sun, 
using $m_{\rm J} = 1.9 \times 10^{30}$ g, $R = 7.8 \times 10^{13}$ cm and 
$\omega = 1.68 \times 10^{-7}$ Hz, we get $P = 5 \times 10^3$ Joules/sec.
This value is tiny, especially when compared to the luminosity of the Sun in 
electromagnetic radiation $P_\odot = 3.9 \times 10^{26}$ Joules/sec. 
If the binary moves along an eccentric orbit the power radiated is~\cite{ecc}
\beq
P = \frac{32}{5}\,\frac{G^4\,\mu^2\,M^2}{a^5\,c^5}\,\frac{1}{(1-e^2)^{7/2}}\,
\left ( 1 + \frac{73}{24}e^2+\frac{37}{96}\,e^4 \right )\,,
\label{HT}
\eeq
where $a$ is the semi-major axis and $e$ the eccentricity. Plugging in Eq.~(\ref{HT}) 
the values for the Hulse-Taylor binary pulsar~\cite{HT75}, 
$a = 1.95 \times 10^{11}$ cm, $m_1 = 1.441 M_\odot$, $m_2 = 1.383 M_\odot$, 
$e = 0.617$, we get 
that the power radiated is $7.35 \times 10^{24}$ Joules/sec, 
which is about $2\%$ of the luminosity of the Sun in electromagnetic radiation.  

The emission of GWs costs energy and to compensate for the loss of energy, 
the radial separation $R$ between the two bodies must decrease. We shall 
now derive how the orbital frequency and GW frequency change in time, 
using Newtonian dynamics and the balance equation  
\beq
\frac{d E_{\rm orbit}}{d t} = - P\,.
\label{eq:84}
\eeq
At Newtonian order, $E_{\rm orbit} = - G\,m_1\,m_2/(2R)$ and 
$\omega^2 = G\,M/R^3$. Thus, $\dot{R} = -2/3\,(R\,\omega)\,
(\dot{\omega}/\omega^2)$. As long as $\dot{\omega}/\omega^2 \ll 1$, the radial 
velocity is smaller than the tangential velocity and the binary's motion  
is well approximated by an adiabatic sequence of quasi-circular orbits.
Equation (\ref{eq:84}) implies that the orbital frequency varies as 
\beq
\frac{\dot{\omega}}{\omega^2}=\frac{96}{5}\,\nu\,\left (\frac{G M\omega}{c^3} \right )^{5/3}\,,
\label{eq}
\eeq
and the GW frequency $f_{\rm GW} = 2 \omega$,  
\beq
\dot{f}_{\rm GW} = \frac{96}{5}\pi^{8/3}\,\left (\frac{G\,{\cal M}}{c^3} \right )^{5/3}\,f_{\rm GW}^{11/3}\,,
\label{eq:85}
\eeq
where ${\cal M} = \mu^{3/5}\,M$ is the so-called {\it chirp} mass. Introducing the time 
to coalescence $\tau = t_{\rm coal} -t$, and integrating Eq.~(\ref{eq:85}), we get
\beq
f_{\rm GW} \simeq 130 \left (\frac{1.21 M_\odot}{\cal M}\right )^{5/8}\, 
\left (\frac{1 {\rm sec}}{\tau} \right )^{3/8}\,{\rm Hz}\,,
\label{eq:86}
\eeq
where $1.21 M_\odot$ is the chirp mass of a NS-NS binary. 
Equation (\ref{eq:86})  predicts coalescence times of 
$ \sim 17 {\rm min}, 2 {\rm sec}, 1 {\rm msec}$, for $f_{\rm GW} \sim 
10, 100 ,10^3$ Hz. Using the above equations, 
it is straightforward  to compute the relation between the radial 
separation and the GW frequency, we find
\beq
R \simeq 300 \left (\frac{M}{2.8 M_\odot} \right )^{1/3}\,
\left (\frac{100\, {\rm Hz}}{f_{\rm GW}} \right )^{2/3}\, {\rm km}\,.
\label{eq:87}
\eeq
Finally, a useful quantity is the number of GW cycles, defined by 
\beq 
{\cal N}_{\rm GW} = \frac{1}{\pi} \int_{t_{\rm in}}^{t_{\rm fin}} 
\omega(t)\,dt = \frac{1}{\pi} \int_{\omega_{\rm in}}^{\omega_{\rm fin}} 
\frac{\omega}{\dot{\omega}}\,d\omega\,.
\label{cycles}
\eeq
Assuming $\omega_{\rm fin} \gg \omega_{\rm in}$, we get 
\beq
{\cal N}_{\rm GW} \simeq 10^4\,\left (\frac{{\cal M}}{1.21 M_\odot} \right )^{-5/3}\,
\left (\frac{f_{\rm in}}{10 {\rm Hz}} \right )^{-5/3}\,.
\label{eq:88}
\eeq

\subsection{Inspiral waveform including post-Newtonian corrections}
\label{sec6.2}

Already in the early developments of Einstein theory, an approximation method  
called post-Newtonian (PN) method was developed by Einstein, Droste and De Sitter.  
This method allowed theorists to draw quickly several  
observational consequences, and within one year of the discovery 
of general relativity, led to the 
predictions of the relativistic advance of the perihelion of planets, 
the gravitational redshift and the deflection of light. 

The PN method involves an expansion around the Newtonian 
limit keeping terms of higher order in the small 
parameter~\cite{TD87,PNreview} 
\beq
\epsilon \sim \frac{v^2}{c^2} \sim \left |h_{\mu \nu} \right | 
\sim \left |\frac{\partial_0 h}{\partial_i h} \right |^2 \sim 
\left |\frac{T^{0i}}{T^{00}}\right | \sim \left |\frac{T^{ij}}{T^{00}}
\right |\,.
\eeq
In order to be able to determine the dynamics of binary systems with a precision 
acceptable for detection, it has been necessary to compute the force determining 
the motion of the two bodies and the amplitude of the gravitational radiation 
with a precision going beyond the quadrupole formula (\ref{eq:62}). 
For nonspinning BHs, the two-body equations of motion and the GW 
flux are currently known through 3.5PN order~\cite{PNnospin}. 
If we restrict 
the discussion to circular orbits, as Eq.~(\ref{eq}) shows, 
there exists a natural {\it adiabatic} parameter $\dot{\omega}/\omega^2 = {\cal O}[(v/c)^5]$. 
Higher-order PN corrections to Eq.~(\ref{eq}) have been 
computed~\cite{PNnospin,PNspin}, yielding ($G=1=c$)
\beq
\frac{\dot{\omega}}{\omega^2} = \frac{96}{5}\,\nu\,v_\omega^{5/3}\,
\sum_{k=0}^7 {\omega}_{(k/2)\mathrm{PN}}\,v_\omega^{k/3}\,
\label{omegadot}
\eeq
where we define $v_\omega \equiv (M\,\omega)^{1/3}$ and 
\bea
{\omega}_{0\mathrm{PN}} &=& 1\,,\\
\label{omegadotSTpn0}
{\omega}_{0.5\mathrm{PN}} &=& 0\,,\\
\label{omegadotSTpn05}
{\omega}_{1\mathrm{PN}} &=& -\frac{743}{336} -\frac{11}{4}\,\nu\,, \\
\label{omegadotSTpn1}
{\omega}_{1.5\mathrm{PN}} &=& 4\pi + \left[-\frac{47}{3}\frac{S_\ell}{M^2}
-\frac{25}{4}\frac{\delta m}{M}\frac{\Sigma_\ell}{M^2}\right]\,,\\
\label{omegadotSTpn15}
{\omega}_{2\mathrm{PN}} &=& 
\frac{34\,103}{18\,144}+\frac{13\,661}{2\,016}\,\nu+\frac{59}{18}\,\nu^2 - 
\frac{1}{48}\, \nu\,\chi_1\chi_2\left[247\,(\hS_1\cdot\hS_2)- \right. \nonumber \\
&& \left. 721\,
(\boldsymbol{\hat{\ell}}\cdot\hS_1)(\boldsymbol{\hat{\ell}}\cdot\hS_2)\right]\,, 
\label{omegadotSTpn2} 
\eea
\bea
{\omega}_{2.5\mathrm{PN}} &=& -\frac{1}{672}\,(4\,159 +15\,876\,\nu)\,\pi + 
\left[
\left(-\frac{31811}{1008}+\frac{5039}{84}\nu\right)\frac{S_\ell}{M^2}+ \right . \nonumber \\
&& \left . \left(-\frac{473}{84}+\frac{1231}{56}\nu\right)\frac{\delta
m}{M}\frac{\Sigma_\ell}{M^2}\right]\,, \\
\label{omegadotSTpn25}
{\omega}_{3\mathrm{PN}} &=& 
\left(\frac{16\,447\,322\,263}{139\,708\,800}-\frac{1\,712}{105}\,\gamma_E+\frac{16}{3}\pi^2\right)+
\left(-\frac{56\,198\,689}{217\,728}+ \right . \nonumber \\
&&\left . \frac{451}{48}\pi^2 \right)\nu 
+\frac{541}{896}\,\nu^2-\frac{5\,605}{2\,592}\,\nu^3
-\frac{856}{105}\log\left[16v^{2}\right]\,,\\
\label{omegadotSTpn3}
{\omega}_{3.5\mathrm{PN}} &=& \left (
-\frac{4\,415}{4\,032}+\frac{358\,675}{6\,048}\,\nu+\frac{91\,495}{1\,512}\,\nu^2
\right )\,\pi\,.
\label{omegadotSTpn35}
\eea
We denote $\boldsymbol{\ell} = \mu \, \vX \times \vV$ the 
Newtonian angular momentum (with $\vX$ and $\vX$ the two-body center-of-mass
radial separation and relative velocity), and $\boldsymbol{\hat{\ell}} = \boldsymbol{\ell} /
|\boldsymbol{\ell}|$; $\vS_1 =\chi_1\,m_1^2\,\hS_1$ and $\vS_2
=\chi_2\,m_2^2\,\hS_2$ are the spins of the two bodies (with
$\hS_{1,2}$ unit vectors, and $0 < \chi_{1,2} < 1$ for BHs) and
\beq
\label{spins}
\mathbf{S} \equiv \mathbf{S}_1 + \mathbf{S}_2\,, \quad \mathbf{\Sigma} \equiv M\left[\frac{\mathbf{S}_2}{m_2} -
\frac{\mathbf{S}_1}{m_1}\right]\,.
\eeq
Finally, $\delta m = m_1-m_2$ and $\gamma_E=0.577\ldots$ is Euler's constant. 

\begin{table*}[t]
\caption{Post-Newtonian contributions to the number of GW
  cycles accumulated from $\omega_\mathrm{in} =
  \pi\times 10\,\mathrm{Hz}$ to $\omega_\mathrm{fin} =
  \omega^\mathrm{ISCO}=1/(6^{3/2}\,M)$ for binaries detectable by
  LIGO and VIRGO. We denote $\kappa_{i} = \hS_i \cdot \boldsymbol{\hat{\ell}}$ and 
   $\xi = \hat{\mathbf{S}}_1\cdot
  \hat{\mathbf{S}}_2$. 
\label{tab:1}}
\begin{center}
{\scriptsize
\begin{tabular}{|l|c|c|}\hline
& \multicolumn{1}{c|}{$(10+10)M_\odot$} &
\multicolumn{1}{c|}{$(1.4+1.4)M_\odot$} \\ \hline\hline 
Newtonian & $601$ & $16034$ \\ 
1PN & $+59.3$ & $+441$\\ 
1.5PN & $-51.4 + 16.0\, \kappa_1\,\chi_1 + 16.0\, \kappa_2\,\chi_2$ 
& $ -211 + 65.7\,\kappa_1\,\chi_1 + 65.7\, \kappa_2\,\chi_2$ \\ 
2PN & $+4.1 - 3.3\, \kappa_1\,\kappa_2\,\chi_1\,\chi_2 + 1.1\, \xi\,\chi_1\,\chi_2$
& $+ 9.9 - 8.0\, \kappa_1\,\kappa_2\,\chi_1\,\chi_2 + 2.8
\,\xi\,\chi_1\,\chi_2$ \\ 
2.5PN & $-7.1 + 5.5\, \kappa_1\,\chi_1 + 5.5\,
\kappa_2\,\chi_2$ & $-11.7 + 9.0\, \kappa_1\,\chi_1 + 9.0\,
\kappa_2\,\chi_2$ \\ 
3PN & $+2.2$ & $+2.6$ \\ 
3.5PN & $-0.8$ & $-0.9$ \\ \hline
\end{tabular}}\end{center}
\end{table*}

It is instructive to compute the relative contribution of the 
PN terms to the total number of GW cycles accumulating 
in the frequency band of LIGO/VIRGO. In Table~\ref{tab:1},  
we list the figures obtained by plugging Eq.~(\ref{omegadot}) 
into Eq.~(\ref{cycles}). As final frequency we use the 
innermost stable circular orbit (ISCO) of a point particle in 
Schwarzschild [$f_{\rm GW}^{\rm ISCO} \simeq 4400/(M/M_\odot)$ Hz]. 

\subsection{Full waveform: inspiral, merger and ring-down}
\label{sec6.3}
\begin{figure}
\begin{center}
\begin{tabular}{cc}
\includegraphics[width=0.6\textwidth]{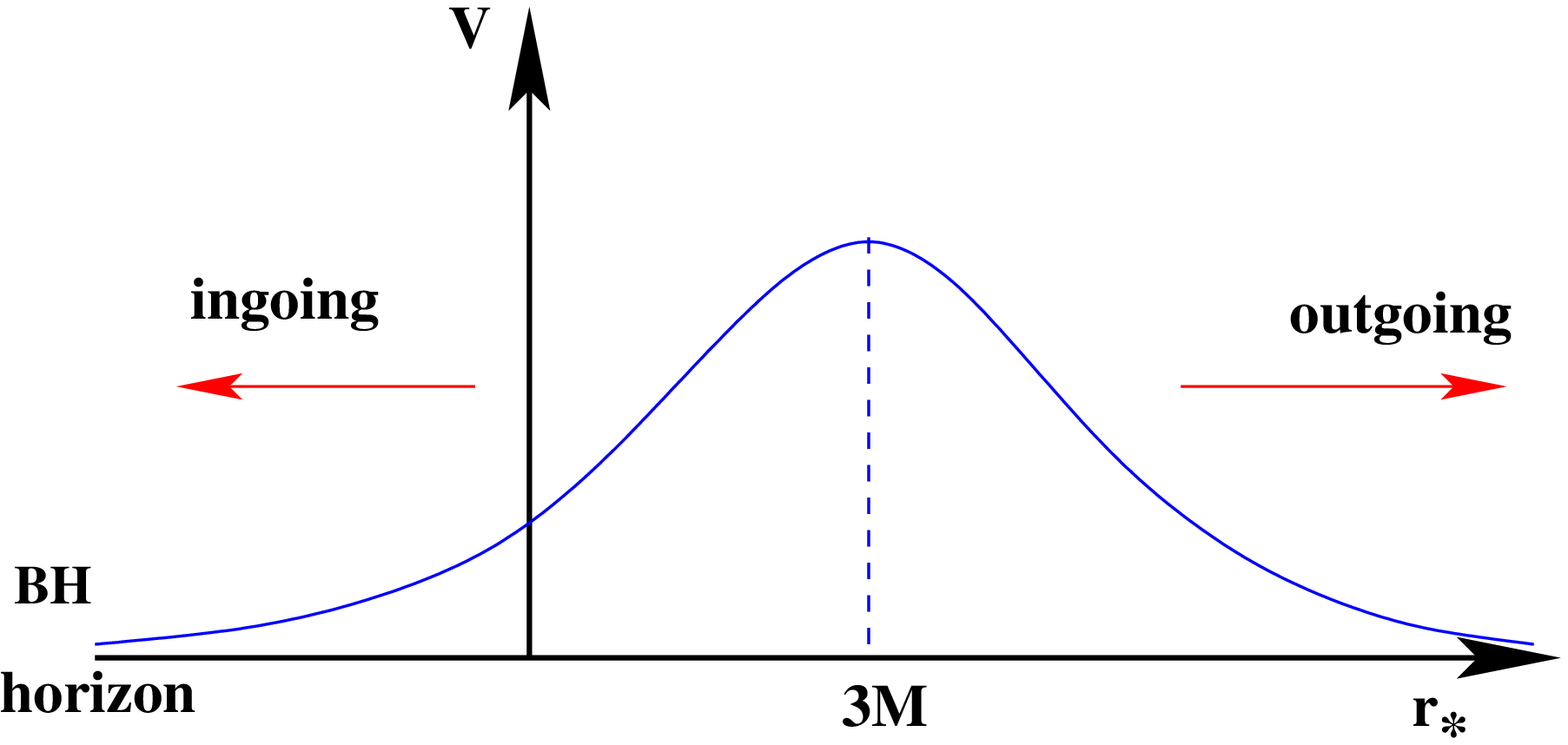} & 
\includegraphics[width=0.3\textwidth]{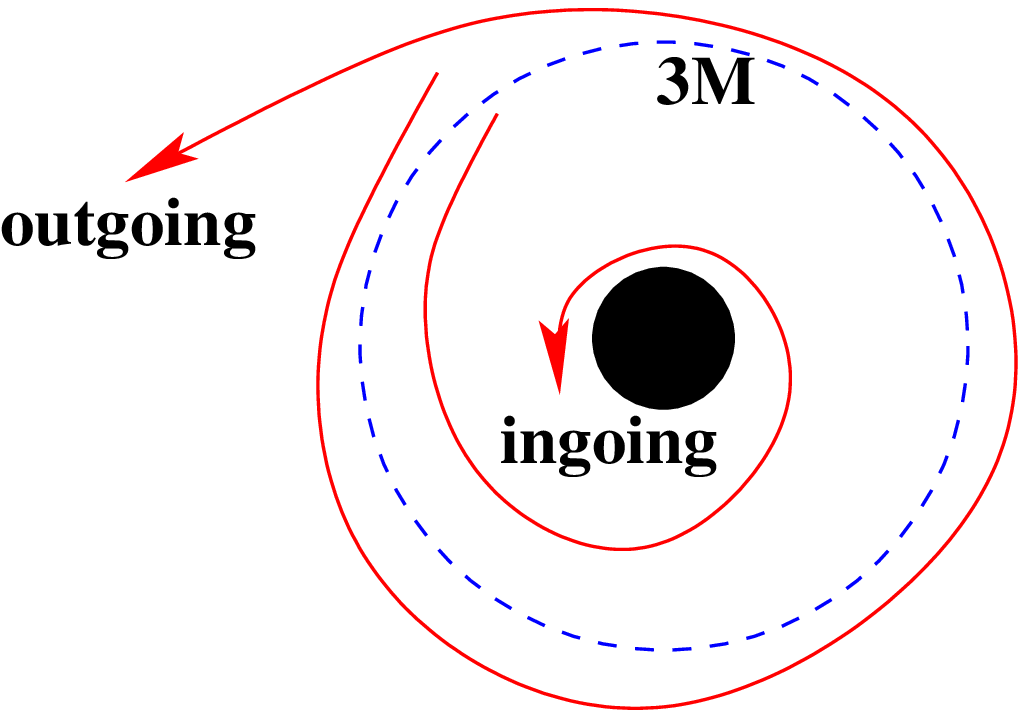}\\
\end{tabular}
\caption{We sketch the curvature potential  
as function of the tortoise coordinate 
$r^*$ associated to metric perturbations of 
a Schwarzschild BH. The potential peaks at the last 
unstable orbit for a massless particle (the light ring). 
Ingoing modes propagate toward the BH horizon, whereas 
outgoing modes propagate away from the source. \label{fig:7}}
\end{center}
\end{figure}
\begin{figure}
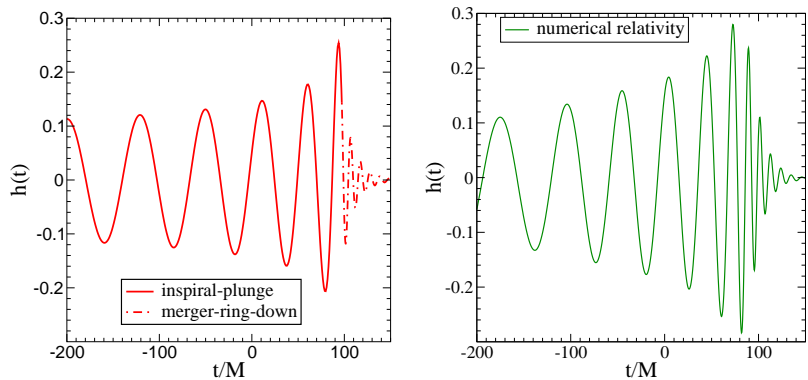

\begin{center}
\begin{tabular}{cc}
\includegraphics[width=0.45\textwidth]{fig5a.eps} &
\includegraphics[width=0.45\textwidth]{fig5b.eps}\\
\end{tabular}
\caption{On the left panel we show the GW signal from an equal-mass 
nonspinning BH binary as predicted at 2.5PN order by Buonanno and Damour (2000) 
in Ref.~\cite{EOB}. 
The merger is assumed almost instantaneous and one QNM is included. 
On the right panel we show the GW signal from an equal-mass 
BH binary with a small spin $\chi_1=\chi_2 = 0.06$ obtained in 
full general relativity by Pretorius~\cite{BCP}
\label{fig:6}}
\end{center}
\end{figure}
\begin{figure}
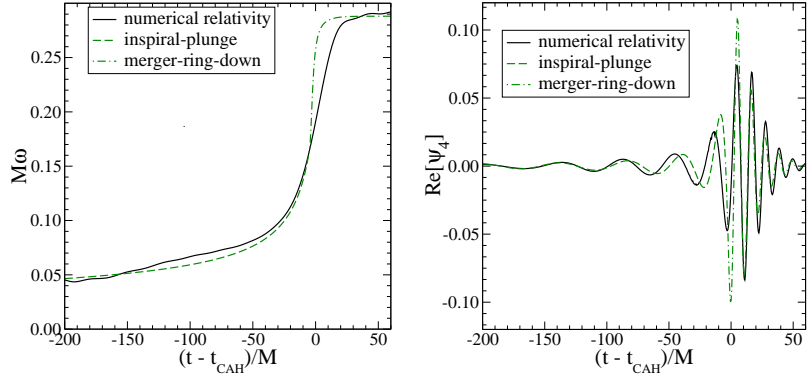

\begin{center}
\begin{tabular}{cc}
\includegraphics[width=0.45\textwidth]{fig6a.eps} &
\includegraphics[width=0.45\textwidth]{fig6b.eps}\\
\end{tabular}
\caption{\label{fig:8} Comparison between inspiral-merger-ring-down 
frequency (left panel) and waveform (right panel) for an equal-mass BH binary with spin $\chi_1=\chi_2 = 0.06$, 
as predicted at 3.5PN order in Ref.~\cite{EOB} and as obtained in 
a numerical simulation by Pretorius~\cite{BCP}. In the analytic 
model the merger is assumed almost instantaneous and three QNMs are 
included~\cite{BCP}. $\Psi_4$ is the Weyl tensor proportional to 
the second derivative of $h$; we denote with $t_{\rm CAH}$ the 
time when the apparent common horizon forms.}
\end{center}
\end{figure}

After the two BHs merge, the system settles down to a Kerr BH 
and emits quasi-normal modes (QNMs), as originally predicted by 
Ref.~\cite{qnm,Press}. This phase is commonly known as 
the ring-down (RD) phase. Since the QNMs have complex frequencies totally 
determined by the BH's mass and spin, the RD waveform is a 
superposition of damped sinusoidals. The inspiral and 
RD waveforms can be computed analytically. What about the 
merger? Since the nonlinearities dominate, 
the merger would be described at {\it best} and {\it utterly} 
through numerical simulations of Einstein equations. However, 
before NR results became available, some 
analytic approaches were proposed. 
In the test-mass limit, $\nu \ll 1$, Refs.~\cite{Davis,Press} realized a long time 
ago that the basic physical reason underlying 
the presence of a universal merger signal was that 
when a test particle falls below $ 3 M$ (the 
unstable light storage ring of Schwarzschild), the GW 
it generates is strongly filtered by the curvature potential 
barrier centered around it (see Fig.~\ref{fig:7}). 
For the equal-mass case $\nu = 1/4$, Price and Pullin~\cite{CLA}  
proposed the so-called close-limit approximation, 
which consists in switching from the two-body description to the 
one-body description (perturbed-BH) close to the light-ring 
location. Based on these observations, 
the effective-one-body (EOB) resummation scheme~\cite{EOB} 
provided a first {\it example} of full 
waveform by (i) resumming the PN Hamiltonian, (ii) 
modeling the merger as a very short (instantaneous) phase 
and (iii) matching the end of the plunge (around the light-ring) 
with the RD phase (see Ref.~\cite{lazarus} where similar 
ideas were developed also in NR). 
The matching was initially done using {\it only} the least 
damped QNM whose mass and spin were determined by the binary BH 
energy and angular momentum at the end of the plunge. 
An example of full waveform is given in Fig.~\ref{fig:6}. 
Today, with the spectacular results in NR, we are in the position of 
assessing the closeness of analytic to numerical waveforms for inspiral, 
merger and RD. In Fig.~\ref{fig:8}, we show some first-order comparisons between 
the EOB-analytic and NR waveforms~\cite{BCP} (see also Ref.~\cite{Goddshort}). 
Similar results for the inspiral phase but using PN theory~\cite{PNnospin,PNspin} 
(without resummation) at 3.5PN order are given in Refs.~\cite{BCP,Goddshort}.
So far, the agreement is qualitatively good, but more accurate simulations, starting 
with the BHs farther apart, are needed to draw robust conclusions.   

Those comparisons are suggesting that it should be possible to 
design purely analytic templates with the full numerics used to guide the
patching together of the inspiral and RD waveforms. 
This is an important avenue to template construction as
eventually hundreds of thousands of waveform 
templates may be needed to extract the signal from 
the noise, an impossible demand for NR alone. 

\subsection{Inspiral templates for data analysis}
\label{sec6.4}
The search for GWs from coalescing binaries 
with laser interferometer GW detectors is based on 
the matched-filtering technique, which requires accurate knowledge of the 
waveform (or template) of the incoming signal. As an example, 
in this section we derive the inspiral GW template in Fourier domain
using the stationary phase approximation (SPA). Those templates are 
currently used to search for inspiraling binary with LIGO/VIRGO/GEO/TAMA 
detectors. Henceforth, we use $G=1=c$. 

The detector response to a GW signal is given by~\cite{DA} 
\beq
h(t) = h_+(t)\,F_+ + h_\times(t)\,F_\times\,,
\label{hresp}
\eeq
\bea
F_+ &=& \frac{1}{2}(1+\cos^2 \Theta)\,\cos 2\Phi\,\cos 2 \Psi - \cos \Theta\,\sin 2 \Phi\,\sin 2\Psi\,, \\
\label{F+}
F_\times &=& \frac{1}{2}(1+\cos^2 \Theta)\,\cos 2\Phi\,\sin 2 \Psi + \cos \Theta\,\sin 2 \Phi\,\cos 2\Psi\,,
\label{Fx}
\eea
being $\Theta, \Phi, \psi$ the angles defining the relative orientation of the binary 
with respect to the detector~\cite{DA}. It is convenient to 
introduce the variables
\bea
\widetilde{F}_+ &\equiv& \frac{(1+\cos^2 \Theta)\,F_+}{[(1+\cos^2 \Theta)^2\,F_+^2 + 4\cos^2 \Theta\,F_\times^2]^{1/2}}\,,\\
\widetilde{F}_\times &\equiv& \frac{4\cos^2 \Theta\,F_\times}{[(1+\cos^2 \Theta)^2\,F_+^2 + 4\cos^2 \Theta\,F_\times^2]^{1/2}}\,.
\eea
Noticing that $\widetilde{F}_+^2 + \widetilde{F}_\times^2 =1$, we can define 
$\cos \xi \equiv \widetilde{F}_+$ and $\sin \xi \equiv \widetilde{F}_\times$, and  
\bea
{\cal A}(\Theta,\Phi,\Psi;\theta) &\equiv & 
[(1+\cos^2 \theta)^2\,F_+^2 + 4\cos^2 \theta\,F_\times^2]^{1/2} \,,\\
\tan \xi(\Theta,\Phi,\Psi;\theta) &\equiv & \frac{4\cos^2 \theta\,F_\times}{(1+\cos^2 \theta)\,F_+}\,.
\eea
The GW signal for an inspiraling binary (\ref{hresp}) can be rewritten in the simpler 
form 
\beq
h(t) = \frac{2 {\cal M}}{r}\,{\cal A}(\Theta,\Phi,\Psi;\theta)\,
[{\cal M}\, \omega (t)]^{2/3}\,\cos[2\phi(t) +2\phi_0-{\xi}]\,.
\eeq
If we are not interested in recovering the binary's orientation with respect 
to the detector (the so-called {\it inverse} problem), we can absorb 
$\xi$ into $\phi_0$, and average over the angles 
$(\Theta,\Phi,\Psi,\theta)$ obtaining~\cite{DA} $\overline{{\cal A}^2} = {16}/{25}$.

Let us now compute the Fourier transform of the GW signal 
\bea
\tilde{h}(f)&=&\int_{-\infty}^{+\infty}\,e^{2\pi i f t}\,h(t)\,dt\,, \nonumber \\
&=&\frac{1}{2}\,\int_{-\infty}^{+\infty}\,dt\,{A(t)}\,
\left [e^{2\pi i f t+ i{\phi_{\rm GW}(t)}}
+ e^{2\pi i f t- i{\phi_{\rm GW}(t)}} \right ]\,,
\label{eqint}
\eea
where $ A(t) = (2 {\cal M}/R)\,\sqrt{{\cal A}^2}\,[{\cal M}\, \omega(t)]^{2/3}$ and 
$\phi_{\rm GW}(t) = 2\phi(t) + 2\Phi_0$. We compute the integral as follows. In Eq.~(\ref{eqint}) 
The dominant contribution comes from the vicinity of the {\it stationary} points in the phase. 
Assuming $f>0$, we pose $\psi(t) \equiv 2\pi\,f\,t - \phi_{\rm GW}$ and 
impose $\left ({d\psi}/{dt}\right )_{t_f} =0$, that is 
$\left ({d \phi_{\rm GW}}/{dt}\right )_{t_f}= 2\pi\,f = 2\pi F(t_f)$. 
Expanding the phase up to quadratic order 
\beq
\psi(t_f) = 2 \pi\,f\,t_f - \phi_{\rm GW}(t_f) - \pi\,\dot{F}(t_f)\,(t-t_f)^2\,,
\eeq
we get
\beq
{\tilde{h}_{\rm SPA}}(f) = \frac{1}{2}\,{\frac{A(t_f)}{\sqrt{\dot{F}(t_f)}}}\,
e^{i [2 \pi\,f\,t_f-{\phi_{\rm GW}(t_f)}]-i\pi/4}\,.
\eeq
To compute $\phi_{\rm GW}(t_f)$ and $\dot{F}(t_f)$, we need to solve 
the following equations
\beq
v^3 = \dot{\phi}_{\rm GW}\,\frac{M}{2}\,, \quad \frac{dE}{dt}(v)=-{\cal F}(v)\,,
\eeq 
where $E$ is the center-of-mass energy and ${\cal F}$ the GW energy flux. A direct 
calculation yields
\bea
t(v) &=& t_{c} + M\,\int_v^{v_{c}}\,dv\,\frac{E'(v)}{{\cal F}(v)}\,, \\
\phi_{\rm GW}(v) &=& \phi_{c} + 2\,\int_v^{v_{c}}\,dv\,v^3\,\frac{E'(v)}{{\cal F}(v)}\,,
\eea
thus, 
\beq
\psi(f) = 2 \pi f\,t_c - \phi_c - \frac{\pi}{4} + 
2 \int_{v}^{v_c} (v_c^3-v^3)\,\frac{E'(v)}{{\cal F}(v)}\,dv\,,
\eeq
Using Eq.~(\ref{eq}), we have  
$\dot{F}(t_f) \equiv {\dot{\omega}}/{\pi} = ({96}/{5})\,({1}/{\pi})\,\nu\,M^{5/3}\,\omega^{11/3}$, 
and we obtain~\cite{spa}
\bea
{\tilde{h}_{\rm SPA}}(f) &=& {{\cal A}_{\rm SPA}}(f)\,e^{i {\psi_{\rm SPA}}(f)}\,,\\
{{\cal A}_{\rm SPA}}(f) &=&
\frac{\sqrt{{\cal A}^2}}{r}\,\frac{1}{\pi^{2/3}}\,\left (\frac{5}{96}\right )^{1/2}\,{\cal M}^{5/6}\, f^{-7/6}\,, \\
\psi_{\rm SPA}(f) &=& 2\pi f t_c-\phi_c-\frac{\pi}{4}
+{\frac{3}{128\,\nu\, v_f^5}}\;\sum_{k=0}^{7}\psi_{(k/2)\mathrm{PN}}\,v_f^k\,,
\label{phaseSPA}
\eea
where we denote $v_f= (\pi M f)^{1/3}$. 
The coefficients $\psi_{(k/2)\mathrm{PN}}$'s, $k=0,\ldots,7,$ (with $N=7$ at 3.5PN order)
in the Fourier phase are given by
\begin{subequations}
\begin{eqnarray}
\psi_{0\mathrm{PN}}&=&1\,,\\
\psi_{0.5\mathrm{PN}}&=&0\,,\\
\psi_{1\mathrm{PN}}&=&\left( \frac{3715}{756} + \frac{55}{9}\nu
\right)\,,\label{eq:alpha2}\\
\psi_{1.5\mathrm{PN}}&=& -16\pi+ 4 \beta\,,\label{eq:alpha3}\\
\psi_{2\mathrm{PN}}&=&\left( \frac{15293365}{508032} + \frac{27145}
{504}\,\nu + \frac{3085}{72}\,\nu^2 \right) - 10 \sigma\,,\\
\psi_{2.5\mathrm{PN}}&=&\pi\left(\frac{38645 }{756} - \frac{65}{9}\nu \right )
\left(1  + 3\log v_f \right)\,,\\
\psi_{3\mathrm{PN}}&=&\left(\frac{11583231236531}{4694215680} - \frac{640\,{\pi
}^2}{3} - \frac{6848\,\gamma_E }{21}\right)+ \nonumber \\
&& \nu \,\left( - \frac{15737765635}{3048192} + \frac{2255\,{\pi }^2}{12} 
\right) +\nonumber\\
&&{\frac{76055}{1728}}\nu^2-{\frac{127825}{1296}}\nu^3-{\frac{6848}{21}}
\log\left(4\;{v}\right)\,,\\
\psi_{3.5\mathrm{PN}}&=&\pi\left(\frac{77096675 }{254016} + \frac{378515
}{1512}\,\nu - \frac{74055}{756}\,\nu^2\right)\,,
\end{eqnarray}\label{eq:alphas}
\end{subequations}
where 
\bea
\beta&=& \sum_{i=1}^2 \left 
(\frac{113}{12}\,\frac{m_i^2}{M^2}+\frac{25}{4}\,\nu \right )\,\chi_i\,\kappa_i \,,\\
\sigma&=& \nu \left 
(\frac{721}{48}\,\chi_1\,\kappa_1\,\chi_2\,\kappa_2 - \frac{247}{48} \xi \right )\,.
\eea
In alternative theories of gravity~\cite{W}, such as Brans-Dicke theory or massive 
graviton theories, the SPA phase (\ref{phaseSPA}) contains the terms  
\beq
\psi_{\rm SPA}^{\rm alt\, th}(f) = 
{\frac{3}{128\,\nu\, v_f^5}}\; \left [- \frac{5 {\cal S}^2}{84 \omega_{\rm BD}}\,v_f^{-2}-
\frac{128}{3}\frac{\pi^2 D\,{\nu\,M}}{\lambda_g^2\,(1+z)}\,v_f^{2} \right ]\,.
\label{alth}
\eeq
The first term in the square brace is the contribution of dipole gravitational radiation
in Brans-Dicke theory. The scalar charge of the $i-$th
body is $\alpha_i=\bar \alpha \hat \alpha_i=\bar \alpha (1-2s_i)$,
where $\bar \alpha^2=1/(2\omega_{\rm BD}+3) \sim (2\omega_{\rm BD})^{-1}$ in
the limit $\omega_{\rm BD}\gg 1$, and $s_i$ is called the {\it sensitivity}
of the $i-$th body (a measure of the self-gravitational binding energy
per unit mass). The coefficient in the dipole term is ${\cal
S}=(\hat \alpha_1-\hat \alpha_2)/2$. The fact that it is dipole
radiation means that it is proportional to $v^{-2}$ compared to the
quadrupole term, but the small size of $\cal S$ and the large current
solar-system bound on $\omega_{\rm BD} > 40,000$ make this a small correction.  
The second term in the square brace of Eq.~(\ref{alth}) is the effect of a massive
graviton, which alters the arrival time of waves of a given frequency,
depending on the size of the graviton Compton wavelength $\lambda_g$
and on a distance quantity $D$, defined in Ref.~\cite{BBW}. It is possible 
to put interesting bounds on $\omega_{\rm BD}$ and $\lambda_g$ using LISA~\cite{W,BBW}. 
\begin{figure}
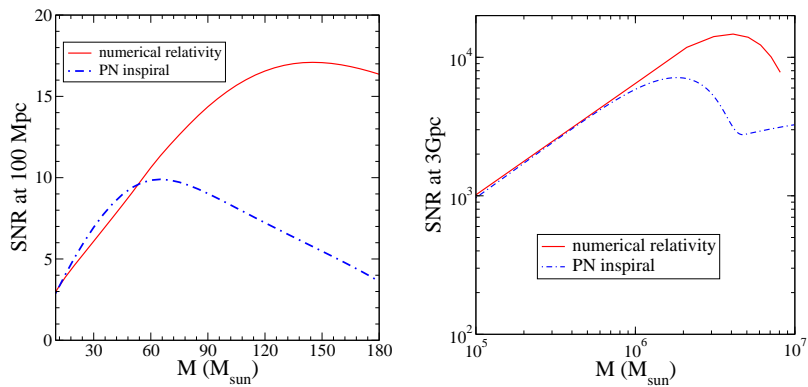

\begin{center}
\begin{tabular}{cc}
\includegraphics[width=0.45\textwidth]{fig7a.eps} &
\includegraphics[width=0.45\textwidth]{fig7b.eps} \\
\end{tabular}
\caption{We show the sky-averaged SNR for an equal-mass 
nonspinning binary when either the PN inspiral waveform 
(terminated at the PN ISCO) or the full NR waveform are included. The left panel uses 
the noise spectral density of LIGO, whereas the right panel 
the noise spectral density of LISA.
\label{fig:5}}
\end{center}
\end{figure}

Finally, the sky-average signal-to-noise ratio (SNR) for SPA waveforms is 
\beq
\sqrt{\overline{\left (\frac{S}{N}\right )^2}} = \frac{1}{r}\,
\frac{1}{\pi^{2/3}}\,\sqrt{\frac{2}{15}}\,{\cal M}^{5/6}\,
\left [ \int_{0}^{f_{\rm fin}} 
\frac{f^{-7/3}}{S_n(f)} \right ]^{1/2}\,.
\eeq
In the left panel of Fig.~\ref{fig:5}, we show the sky average SNRs versus total mass, for an equal-mass 
nonspinning binary at 100 Mpc using the noise-spectral density of LIGO. 
Astrophysical observations and theoretical 
predictions suggest that stellar mass BHs have a total mass ranging between 
$6 \mbox{--} 30 M_\odot$. If binary BHs of larger total mass exist,
they could be detected by LIGO with high SNR.
In the right panel of Fig.~\ref{fig:5} we show the average SNR for one Michelson LISA configuration versus 
the total (redshifted) mass for an equal-mass nonspinning binary at 3 Gpc. The dip in the plot is due to 
the white-dwarf binary confusion noise. Due to the inclusion of merger and ring-down phases, 
the SNR increases considerably for total masses larger than $2 \times 10^6 M_\odot$. 

\section{Other  astrophysical sources}
\label{sec7}

\subsection{Pulsars}
\label{sec7.1}

The production of GWs from a rotating body is of considerable importance, in 
particular for application to isolated pulsars. 
For simplicity we shall examine the case the rigid body rotates around one of its 
principal axis. 

Let us denote by $(X'_1,X'_2,X'_3)$ the coordinates in the reference 
frame attached to the body, the so-called {\it body frame}, and introduce 
a {\it fixed frame} with coordinates $(X_1,X_2,X_3)$, oriented 
such that $X'_3 = X_3$. In both frames the origin of the axes coincides 
with the center-of-mass of the body. The two frames are related by the 
time-dependent rotation matrix
\beq
{\cal R}_{ij} = \left ( \begin{array}{ccc}
\cos \omega\,t & \sin \omega\,t & 0 \\
 - \sin \omega \,t &\cos \omega\,t & 0 \\
 0 & 0&  1 \\
\end{array}
\right )\,, 
\eeq
where $\omega$ is the rotation frequency. The inertia tensor of a rigid body 
is defined by 
\beq
I^{ij} = \int d^3 X\,\rho(\mathbf{X})\,(R^2\,\delta^{ij} - X^i\,X^j)\,,
\eeq
$\rho$ being the mass density. Since any hermitian matrix can be 
diagonalized, there exists an orthogonal frame where $I_{ij}$ is diagonal. 
The eigenvalues are called principal moments of inertia. The frame 
where the inertia tensor is diagonal is the body frame. We denote 
by $I'_{ij} = {\rm diag}(I_1,I_2,I_3)$ the inertia tensor 
in the $(X'_1,X'_2,X'_3)$ coordinate system, and by $I_{ij}$ 
its components in the $(X_1,X_2,X_3)$ frame. Using the 
relation $I = {\cal R}^t\,I'\,{\cal R}$, it is straightforward to derive
\bea
I_{11} &=& 1 + \frac{1}{2}\,(I_1 - I_2)\,\cos (2 \omega\,t)\,,\\
I_{12} &=& \frac{1}{2}\,(I_1 - I_2)\,\sin (2 \omega\,t)\,,
\eea
\bea
I_{22} &=& 1-\frac{1}{2}\,(I_1 - I_2)\,\cos (2 \omega\,t)\,,\\
I_{33} &=& I_3\,,\\
I_{13} &=& I_{23}=0\,.
\eea
To get the leading-order GW signal, we need to compute 
the second-time derivative of the quadrupole tensor 
$M_{ij} = \int d^3 X\,\rho(\mathbf{X})\,X^i\,X^j$, i.e. 
$\ddot{M}_{ij} = - \ddot{I}_{ij}$. We obtain
\bea
M_{11} &=& - \frac{1}{2}\,(I_1-I_2)\,\cos (2 \omega \,t)\,,\\
M_{12} &=& - \frac{1}{2}\,(I_1-I_2)\,\sin (2 \omega \,t)\,,\\
M_{22} &=& \frac{1}{2}\,(I_1-I_2)\,\cos (2 \omega \,t)\,.
\eea
Thus, if the body rotates around the principal axis $X'_3$, there is 
a time-varying second mass moment only if $I_1 \neq I_2$. Plugging in Eqs.~(\ref{eq:62}), 
(\ref{eq:63}) the above expressions, yields 
\bea
h_+ &=& \frac{1}{r}\,\frac{4 G\,\omega^2}{c^4}\,(I_1-I_2)\,\frac{(1+\cos \theta)}{2}\,\cos (2 \omega t)\,,\\
h_\times &=& \frac{1}{r}\,\frac{4 G\,\omega^2}{c^4}\,(I_1-I_2)\,\cos \theta\,\sin (2 \omega t)\,.
\eea
The GW signal is emitted at twice the pulsar rotation frequency. It is common to define 
the ellipticity $\epsilon = (I_1-I_2)/I_3$ and $h_0 = (4 \pi^2G/c^4) (I_3 f^2_{\rm GW}/r)\epsilon$.  
The value of the ellipticity 
depends on the NS properties, in particular the maximum strain that can be 
supported by its crust. Pulsars are thought to form in supernova explosions. 
The outer layers of the star crystallize as the newborn pulsar cools by neutrino 
emission. Anisotropic stresses during this phase could lead to 
values $\epsilon\, \laq\, 10^{-6}$~\cite{CT02}, although with exotic equation of 
state $\epsilon \simeq 10^{-5} \mbox{--} 10^{-4}$~\cite{BO}. Plugging numbers we find
\beq
h_0 \simeq 10^{-25}\,\left (\frac{\epsilon}{10^{-6}} \right )\,\left (\frac{I_3}{10^{45}\,{\rm g\,cm^2}}\right )\,
\left ( \frac{10 {\rm kpc}}{r}\right )\,\left ( \frac{f_{\rm GW}}{1 {\rm kHz}}\right )^2\,.
\eeq
Using Eq.~(\ref{eq:73}), we can compute the total power radiated 
\beq
P= \frac{32}{5}\frac{G}{c^5}\,\epsilon^3\,I_3^2\,\omega^6\,.
\eeq
Due to GW emission, the rotational energy of the star decreases as 
\beq
\frac{d E}{dt} = -\frac{32}{5}\, \frac{G}{c^5}\,\epsilon^3\,I_3^2\,\omega^6\,.
\eeq
Since the rotational energy of a star rotating around its 
principal axis is 
\beq
E = \frac{1}{2}\,I_3\,\omega^2\,,
\eeq
if the GW emission were the 
dominant mechanism for the loss of rotational energy, the pulsar 
frequency should decrease as 
\beq
\dot{\omega} = - \frac{32}{5}\,\frac{G}{c^5}\,\epsilon^2\,I_3\,\omega^5\,.
\eeq
From electromagnetic observations one finds instead $\dot{\omega} \propto 
- \omega^n$, where $n \simeq 2 \mbox{--} 3$. Thus, the GW emission is not 
the major mechanism of energy loss for a rotating pulsar. 

If the rotation axis does not coincide with a principal axis, the 
pulsar motion is a combination of rotation around the principal axis 
and precession. New features appear in the GW signal, as discussed 
in detail in Ref.~\cite{pulsars}.

The detection of continuous, monochromatic frequency waves is achieved 
by constructing power spectrum estimators and searching for statistically 
significant peaks at fixed frequencies for very long time. If $T$ is the observation time, 
the signal-to-noise ratio grows like $\sqrt{T}$. The detection is complicated 
by the fact that the signal received at the detector is not perfectly 
monochromatic due to the Earth's motion. Because of Doppler shifts in frequency, 
the spectral lines of fixed frequency sources spread power into many Fourier 
bins about the observed frequency. Given the possibility that the strongest 
sources of continuous GWs may be electromagnetically invisible or 
previously undiscovered, an all sky, all frequency search for such unknown 
sources is very important, though computationally very expensive
~\cite{algorithms}. 
 
\subsection{Supernovae}
\label{sec7.2}

Supernovae are triggered by the violent collapse of a stellar core which forms 
a NS or BH. The core collapse proceeds extremely fast, lasting 
less than a second and the dense fluid undergoes motions with 
relativistic speeds. Small deviations from spherical 
symmetry during this phase can generate GWs. 

From electromagnetic observations we know that stars with 
mass $M > 8 M_\odot$ end their evolution in core collapse 
and that $90 \%$ of them have mass between $8 \mbox{--} 20 M_\odot$.
During the collapse most of the material is ejected and if the 
progenitor star has a mass $M \leq 20 M_\odot$, it leaves behind 
a NS. If $M \geq 20 M_\odot$, more than $10 \%$ of the material 
falls back and pushes the proto-NS above the maximum NS mass, 
leading to a BH. If the progenitor star 
has a mass $M \geq 40 M_\odot$, no supernovae takes place, the 
star collapses directly to a BH. Numerical simulations~\cite{SN} 
have predicted strains on the order of  
\beq
h = 6 \times 10^{-17}\,\sqrt{\eta_{\rm eff}}\, 
\left (\frac{M}{M_\odot} \right )^{1/2}\,
\left (\frac{10 {\rm kpc}}{r} \right )^{1/2}\,
\left (\frac{1 {\rm kHz}}{f} \right )\,
\left (\frac{10 {\rm msec}}{\tau_{\rm collapse}} \right )^{1/2}\,,
\eeq
where ${\Delta E_{\rm GW}} = \eta_{\rm eff}\,M\,c^2$, $\eta_{\rm eff}\sim 10^{-9}$. 
Reference~\cite{SN} pointed out that GWs could also be produced by neutrino 
emission during the supernovae explosion. The GW signal would  
extend toward lower frequencies $\sim 10$ Hz. Moreover, the superposition 
of independent GW signals from cosmological supernovae 
may give rise to a stochastic GW background. While the estimates remain
  uncertain within several orders of magnitude, this background
  may become detectable by second-generation space-based
  interferometers operating around $\sim 0.1$ Hz~\cite{BSigl}.
 
Note that after a supernovae explosion or a collapsar a 
significant amount of the ejected material can fall back, subsequently 
spinning and heating the NS or BH. Quasi-normal modes can be excited. 
There is also the possibility that the collapsed material 
might fragment into clumps which orbit for some cycles 
like a a binary system or form bar-like structures, which 
also produced GW signals~\cite{mes}. 

\section{Cosmological sources}
\label{sec8}

In this section we want to review stochastic GW backgrounds.  
Depending on its origin, the stochastic background can be broadly divided 
into two classes: the astrophysically generated background due to the incoherent superposition of 
gravitational radiation emitted by large populations of astrophysical sources 
that cannot be resolved individually~\cite{FP}, and the primordial GW background 
generated by processes taking place in the early stages of the Universe. 
A primordial component of such background is especially interesting, since it 
would carry unique information about the state of the primordial Universe. 

The energy and spectral content of such radiation is encoded in the 
spectrum, defined as follows 
\beq
\Omega_{\rm GW}=\frac{1}{\rho_c}\,f\tilde{\rho}_{\rm GW}(f)\,,
\label{eq:omegaf}
\eeq
where $f$ is the frequency, $\rho_c$ is the critical energy density of the 
Universe ($\rho_c=3H^2_0/8\pi G$)  and $\tilde{\rho}_{\rm GW}$ 
is the GWs energy density per unit frequency, i.e. 
\beq
\rho_{\rm GW}=\int_0^{\infty}df\,\tilde{\rho}_{\rm GW}(f)\,.
\eeq
Before discussing the mechanisms that might have generated a 
primordial GW background, we review the main phenomenological 
bounds. 

\subsection{Phenomenological bounds}
\label{sec8.1}

The theory of big-bang nucleosynthesis (BBN) 
predicts rather successfully the primordial abundances 
of light elements. If at $t_{\rm BBN}$, the 
contribution of primordial GWs (or any other extra energy component) 
to the total energy density were to large, 
then the expansion rate of the Universe $H$ will be too large and the 
freeze-out temperature which determines the relative abundance of 
neutrons and protons will be too high. As a consequence, neutrons 
will be more available and light elements will be overproduced spoiling 
BBN predictions. Detailed calculations provide the following bound on the 
energy density~\cite{CST97} 
\beq
\int d \ln f\,h_0^2\,\Omega_{\rm GW}(f) \leq 5.6 \times 10^{-6}\,(N_\nu - 3)\,,
\label{eq:89}
\eeq
where $N_\nu$ is the {\it effective number of neutrino species} at $t_{\rm BBN}$~\cite{M00}. 
The above bound extends to frequency (today) greater than $\sim 10^{-10}$ Hz.
More recently, Ref.~\cite{TPK} derived a similar bound by constraining  
the primordial GW energy density at the time of decoupling $t_{\rm dec}$. 
The latter bound extends to lower frequency 
(today) $\sim 10^{-15}$ Hz, being determined by the comoving horizon size 
at the time of decoupling.   

Another important bound is the so-called COBE bound, which comes 
from the measurement of temperature fluctuations in the CMB produced by 
the Sachs-Wolfe effect. If $\delta T$ is the temperature fluctuation 
\beq
\Omega_{\rm GW}(f) \leq \left ( \frac{H_0}{f} \right )^2\, \left (\frac{\delta T}{T}\right)^2 
\quad 3 \times 10^{-18} {\rm Hz} < f < 10^{-16} {\rm Hz}\,.
\label{eq:90}
\eeq
The lower frequency is fixed by demanding that the fluctuations should be inside the 
Hubble radius today, whereas the higher frequency by imposing that fluctuations 
were outside the Hubble radius at the time of the last scattering 
surface. Detailed analysis give~\cite{AK94} 
\beq
h_0^2\,\Omega_{\rm GW}(f) \leq 7 \times 10^{-11}\,\left (\frac{H_0}{f}\right )^2\,.
\label{eq:91}
\eeq
GWs could saturate this bound if the contribution from scalar perturbations is 
subdominant and this depends on the specific inflationary model.

The very accurate timing of millisec pulsars can constrain $\Omega_{\rm GW}$. 
If a GW passes between us and the pulsar, the time of arrival of the pulse 
is Doppler shifted. Many years of observation yield to the bound~\cite{TD96} (see also 
Ref.~\cite{jenetetal}) 
\beq
h_0^2\,\Omega_{\rm GW} \laq 4.8 \times 10^{-9}\,\left ( \frac{f}{f_*}\right)^2 
\quad f > f_* = 4.4 \times 10^{-9} {\rm Hz}\,.
\label{eq:92}
\eeq

\subsection{Gravitational waves produced by causal mechanisms}
\label{sec8.2}

Two features determine the typical frequency of GWs of cosmological 
origin produced by causal mechanism: (i) the dynamics, which is 
model dependent and (ii) the kinematics, i.e. the redshift from the 
production time. Let us assume that a graviton is produced with 
frequency $f_*$ at time $t_*$ during matter or radiation era. 
What is the frequency today? We have
\beq
f = f_*\,\frac{a_*}{a_0}\,,
\label{eq:93}
\eeq
assuming that the Universe evolved adiabatically, so $g_S(T_*)\,T_*^3\,a_*^3 = 
g_S(T_0)\,T_0^3\,a_0^3$, using $T_0 = 2.73$ K and $g_S(T_0) = 3.91$, where 
$g_S(T_*)$ is the number of degrees of freedom at temperature $T_*$, we get 
\beq
f \simeq 10^{-13}\,f_*\,\left ( \frac{100}{g_{S *}}\right )^{1/3}\,\left (\frac{1 {\rm GeV}}{T_*} 
\right )\,.
\label{eq:94}
\eeq
What is $f_*$? Since the size of the Hubble radius is the scale beyond 
which causal microphysics cannot operate, we can say that from causality 
considerations, the characteristic wavelength of gravitons produced at 
time $t_*$ is $H_*^{-1}$ or smaller. Thus, we set 
\beq
\lambda_* = \frac{\epsilon}{H_*} \quad \epsilon \, \laq \, 1\,.
\label{eq:95}
\eeq
If the GW signal is produced during the radiation era, 
\beq
H_*^2 = \frac{8 \pi G}{3}\,\rho = 8 \pi^3\,g_*\,T_*^4\,\frac{1}{90}\,\frac{1}{M^2_{\rm Pl}}\,,
\label{eq:96}
\eeq
and we find 
\beq
f \simeq 10^{-7}\, \frac{1}{\epsilon}\,\left (\frac{T_*}{1 {\rm GeV}}\right )\,\left (
\frac{g_*}{100}\right )^{1/6}\,{\rm Hz}\,.
\label{eq:97}
\eeq
From the above equation, we obtain that millisec pulsars can probe physics at the $\sim$ MeV 
scale, LISA at the $\sim$ TeV scale, ground-based detectors at $10^3 \mbox{--} 10^{6}$ TeV and detectors 
in the GHz bandwidth~\cite{MWC} at GUT or Planck scale. 

As an application of GWs produced by causal mechanisms, let us consider GW signals from first-order phase transitions. 
In the history of the Universe several phase transitions could have happened. The 
quantum-chromodynamics (QCD)  
phase transition takes place at $T \sim 150$ MeV. Around $T \sim 100$ GeV the 
electroweak (EW) phase transition happens (and $SU(2) \times U(1)$ breaks to 
$U_{\rm em}(1)$ through the Higgs mechanism). 
Let us assume that $V(\phi,T)$ is the potential associated to the 
phase transition, where $\phi$ is the order (field) parameter. 
As the Universe cools down, the {\it true} and {\it false} vacuum 
are separated by a potential barrier. Through quantum tunneling,  
bubbles of true vacuum can nucleate. The difference of energy between the 
true and false vacua is converted in kinetic energy (speed) of 
the bubble's wall. In order to start expanding, the bubbles must have 
the right size, so that the volume energy overcomes the shrinking 
effects of the surface tension. The relevant parameter is the nucleation 
rate $\Gamma = \Gamma_0\,e^{\beta \,t}$. If $\Gamma$ is large enough, 
the bubbles can collide within the Hubble radius, and being the collision 
nonsymmetric, they produce GWs~\cite{WH}. The parameter $\beta$ fixes the frequency 
at the time of production. 

So, we can write for the parameter $\epsilon$ 
defined in Eq.~(\ref{eq:95}), $\epsilon = H_*/f_*=H_*/\beta$ and 
the peak of the GW spectrum occurs at~\cite{BubbleColl,AMNR01,KMK02} 
\beq
f_{\rm peak} \simeq 10^{-8}\,\left (\frac{\beta}{H_*}\right )\,
\left (\frac{T_*}{1 {\rm GeV}}\right )\,\left (\frac{g_*}{100}\right )^{1/6}\, {\rm Hz}\,,
\label{eq:98}
\eeq
where $T_*$ is the temperature at which the probability that a bubble is nucleated within 
the  horizon  size is ${\cal O}(1)$. 

In the case of EW phase transitions, we have $\beta/H_* \simeq 10^2 \mbox{--} 10^3$ and 
$T_* \sim 10^2$ GeV, thus $f_{\rm peak} \simeq 10^{-4} \mbox{--} 10^{-3}$ Hz, which is in 
the frequency band of LISA. The GW spectrum can be computed semianalytically, it 
reads~\cite{BubbleColl,AMNR01,KMK02} 
\beq
h_0^2\,\Omega_{\rm GW} \simeq 10^{-6}\,k^2\,\frac{\alpha^2}{(1+\alpha)^2}\,
\frac{v_b^3}{(0.24 + v_b^3)}\,\left (\frac{H_*}{\beta}\right)^2\,\left (\frac{100}{g_*}\right)^{1/3}\,,
\label{eq:99}
\eeq
where $\alpha$ is the ratio between the false
vacuum energy density and the energy density of the radiation at the
transition temperature $T_*$; $\kappa$ quantifies the fraction of latent heat that is transformed 
into bubble-wall kinetic energy and $v_b$ is the bubble expansion velocity. 

Nonperturbative calculations obtained using lattice field theory have shown that there 
is no first-order phase transition in the standard model if the Higgs mass is larger 
than $M_W$. In minimal supersymmetric standard model, if Higgs mass is $\sim 
110 \mbox{--} 115$ GeV, the transition is first-order but $h_0\,\Omega_{\rm GW} \sim 10^{-19}$ 
at $f \sim 10 $ mHz. In next-to-minimal supersymmetric standard models, there exist 
regions of the parameter space in which $h_0\,\Omega_{\rm GW} \sim 10^{-15}\mbox{--} 10^{-10}$ 
at $f \sim 10 $ mHz. Note that for frequencies 
$10^{-4}\mbox{--}3 \times 10^{-3}$ Hz the stochastic GW background 
from white-dwarf binaries could {\it hide} part of the GW spectrum from first-order 
phase transitions. More recently, Ref.~\cite{GS} pointed out that 
new models of EW symmetry breaking that have been 
proposed have typically a Higgs potential different  
from the one in the Standard Model. Those potentials 
could lead to a stronger first-order phase transition, 
thus to a more promising GW signature in the milliHz frequency 
range. 

A stochastic GW background could be also produced during a first-order
phase transition from turbulent (anisotropic) eddies generated  
in the background fluid during the fast expansion and collision 
of the true-vacuum bubbles~\cite{KKT94,AMNR01,KMK02}. In the 
next-to-leading supersymmetric standard model there exist regions 
of the parameter space where~\cite{AMNR01} $h_0^2\Omega_{\rm GW} \sim 10^{-10}$ with 
peak frequency in the mHz. Reference~\cite{DGN02} 
evaluated the stochastic GW background generated by 
cosmic turbulence before neutrino decoupling, i.e. much later than 
EW phase transition, and at the end of a first-order phase transition 
if magnetic fields also affect the turbulent energy spectrum. 
The observational perspectives of those scenarios are 
promising for LISA.  

Long time ago Turner and Wilczek~\cite{TW90} pointed out that if inflation 
ends with bubble collisions, as in extended inflation, 
the GW spectrum produced has a peak in the frequency range 
of ground-based detectors. Subsequent analyses have shown that
in two-field inflationary models where a field 
performs the first-order phase transition and a second field 
provides the inflationary slow rolling (so-called first-order 
or false vacuum inflation~\cite{FOI}), if the phase transition 
occurs well before the end of inflation, a GW spectrum peaked 
around $10 \mbox{--} 10^3$ Hz, can be produced~\cite{BAFO97}, with an amplitude 
large enough, depending on the number of e-foldings left after the 
phase transition, to be detectable by ground-based 
interferometers. A successful detection of such a spectrum 
will allow to distinguish between inflation and other cosmological 
phase transitions, like QCD or EW, which have a different peak frequency. 

Another mechanism that could have produced GWs 
is parametric amplification after preheating~\cite{KT97}. During 
this phase classical fluctuations produced by the oscillations of 
the inflaton field $\phi$ can interact back, via parametric resonance, 
on the oscillating background producing GWs. In the model where 
the inflaton potential contains also the 
interaction term $\sim \phi^2\,\chi^2$, $\chi$ being a scalar field,  
the authors of Ref.~\cite{KT97} estimated 
$\Omega_{\rm GW} \sim 10^{-12}$ at $f_{\rm min} \sim 10^5$ Hz, while in 
pure chaotic inflation $\Omega_{\rm GW} \,\leq\, 10^{-11}$ at 
$f_{\rm min} \sim 10^4$ Hz. [See Fig. 3 in Ref.~\cite{KT97} 
for the GW spectrum in the range $10^6\mbox{--}10^8$ Hz and 
Refs.~\cite{newprehe} for a recent reanalysis.]
Unfortunately, the predictions lay in the frequency range 
where no GW detectors exist, but some proposals 
have been made~\cite{MWC}

\subsection{Gravitational waves produced by cosmic and fundamental strings}
\label{sec8.3}

Topological defects could have been produced during symmetry-breaking phase transitions 
in the early Universe. Since the 80s they received significant attention as possible candidates 
for seeding structure formation. Recently, more accurate observations 
of CMB inhomogeneities on smaller angular scales and compatibility with the 
density fluctuation spectrum on scales of 100 $h_0^{-1}$ Mpc, restrict the contribution 
of topological defects to less than $10 \%$. 

Cosmic strings are characterized by a single dimensional scale, the mass-per-unit length 
$\mu$. The string length is defined as the energy of the loop divided by $\mu$. 
Cosmic string are stable against all types of decay, except from the 
emission of GWs. Let us assume that a network of cosmic strings did form  
during the evolution of the Universe. In this network the 
only relevant scale is the Hubble length. Small loops 
(smaller than Hubble radius length) oscillate, 
emit GWs and disappear, but they are all the time replaced by small 
loops broken off very long loops (longer than Hubble radius). The wavelength 
of the GW is determined by the length of the loop, and  
since in the network there are loops of all sizes, 
the GW spectrum is (almost) flat in a large frequency band, 
extending from $f\sim 10^{-8}$ Hz to $f\sim 10^{10}$ Hz.  

In 2001, Damour and Vilenkin~\cite{DV01} (see also Ref.~\cite{BHV}), 
worked out that strong bursts of GWs could be produced 
at cusps (where the string reaches 
a speed very close to the light speed) and at kinks along the string 
loop. As a consequence of these bursts the GW background emitted by a 
string network is strongly non-Gaussian. The most interesting feature of these GW bursts is 
that they could be detectable for a large range of values of 
$G \mu$, larger than the usually considered search for the 
Gaussian spectrum. GW bursts can be also produced by fundamental 
strings, as pointed out in Ref.~\cite{fundstring}. For a 
detailed analysis of the prospects of detecting the stochastic  
GW background and the GW bursts with ground and space-based detectors, and 
msec pulsars see Refs.~\cite{XSetal,jenetetal}. 

A GW burst emitted at the cusp of cosmic or fundamental strings can be detected 
using matched filtering. In Fourier domain the signal is~\cite{XSetal}
\beq
h(f) = A |f|^{-4/3}\,\Theta(f_h - f)\,\Theta(f - f_e)\,,
\label{eq:100}
\eeq
where $A \sim (G \mu L^{2/3})/r$, $f_e$ is determined by the size 
$L$ of the feature that produces the cusp, but also by the 
low-frequency cutoff frequency of the detector and $f_h \sim 
2/(\theta^3 L)$, $\theta$ being the angle between the line of sight and the 
direction of the cusp.

\subsection{Gravitational waves produced during inflation}
\label{sec8.4}

The amplification of quantum vacuum fluctuations is a common mechanism in quantum field 
theory in curved space time\cite{BD}. In the 70s Grishchuk and Starobinsky~\cite{russians} 
applied it to cosmology, predicting a stochastic GW background which today would span a very large frequency band
$10^{-17} \mbox{--}10^{10}$ Hz. Henceforth, we shall compute the GW spectrum using 
semiclassical arguments, and refer the reader to Refs.~\cite{old,new} 
for more detailed computations. 

The background field dynamics is described by the action 
\beq
S = \frac{1}{16 \pi G}\,\int d^4 x\,\sqrt{-g}\,R + S_{\rm m}\,.
\label{eq:101}
\eeq
We assume an isotropic and spatially homogeneous Friedmann-
Lamaitre-Robertson-Walker metric with scale factor $a$,
\beq
ds^2 = -dt^2 + a^2\,d \mathbf{x}^2 = g_{\mu \nu}\,dx^\mu\,dx^\nu\,.
\label{eq:102}
\eeq
For what we have learned in previous lectures, we can derive the free-linearized 
wave equations for the TT metric perturbations $\delta g_{\mu \nu} = h_{\mu \nu}$, with 
$h_{\mu 0} =0, h_\mu^\mu=0, h^\mu_{\nu;\mu}=0$ and $\delta T_\mu^\nu = 0$, 
obtaining 
\beq
\Box h_i^j = \frac{1}{\sqrt{-g}} \,\partial_\mu ( \sqrt{-g}\,g^{\mu \nu}\,\partial_\nu) h_i^j=0\,,
\label{eq:103}
\eeq
where we disregard any anisotropic stresses. Introducing the conformal time $\eta$, with 
$d \eta = dt/a(t)$, we can write 
\beq
h_i^j(\eta,\mathbf{x}) = \sqrt{8 \pi G}\,\sum_{A = +,\times} \sum_{\mathbf{k}} h_{\mathbf{k}}^A(\eta)\,
e^{i \mathbf{k}\cdot \mathbf{x}}\,e^{A\,j}_{i}(\mathbf{n})\,,
\label{eq:104}
\eeq
$e^{A\,j}_{i}(\mathbf{n})$ being the polarization tensors. 
Since we assume isotropic and spatially homogeneous metric perturbations 
$h_{\mathbf{k}} = h_{k}$, and each polarization mode satisfies the equation 
\beq
h_k''(\eta) + 2 \frac{a'}{a}\,h'_k(\eta) + k^2\,h_k(\eta) = 0\,,
\label{eq:105}
\eeq
where we denote with a prime the derivative with respect to 
the conformal time. By introducing the canonical variable 
$\psi_k(\eta) = a\,h_k(\eta)$ we can recast Eq.~(\ref{eq:105}) in the 
much simpler form
\beq
\psi_k''(\eta) + \left [k^2 - U(\eta) \right ] \psi_k = 0\,, \quad \quad U(\eta) = \frac{a''}{a}\,,
\label{eq:106}
\eeq
which is the equation of an harmonic oscillator in a time-depedent potential $U(\eta)$. 
We want to solve the above equation and study the properties of the solutions. 

For simplicity, we consider a De Sitter inflationary era, $a = - 1/(H_{\rm DS}\,\eta)$ 
and $a''/a = 2/\eta^2$, and make the crude assumption that the De Sitter era is 
followed instantaneously by the radiation era, $a(\eta) = (\eta - 2 \eta_*)/(H_{\rm DS}\,\eta_*^2)$ and 
$a''(\eta) =0$. 

If $k^2 \gg U(\eta)$, i.e. $k \eta \gg 1$ or $a/k \ll |a \eta| = H_{\rm DS}^{-1}$, the 
mode is inside the Hubble radius or (in jargon) is over the potential barrier $U(\eta)$ and 
the (positive frequency) solution reads
\beq
\psi_k \sim \frac{1}{\sqrt{2 k}}\,e^{- ik\eta} \quad 
\Rightarrow \quad h_k \sim \frac{1}{2 k}\,\frac{1}{a}\,e^{- ik \eta}\,, 
\label{eq:107}
\eeq
thus $h_k$ decreases while inside the Hubble radius. If $k^2 \ll U(\eta)$, i.e. 
$ k \eta \ll 1$,  $a/k \gg |a \eta| = H_{\rm DS}^{-1}$, the mode 
is outside the Hubble radius or (in jargon) under the potential barrier. In this 
case the solution reads
\beq
\psi_k \sim a \left [ A_k + B_k \, \int \frac{d \eta}{a^2(\eta)} \right ] 
\quad \Rightarrow \quad h_k \sim A_k + B_k\,\int \frac{d \eta}{a^2(\eta)}\,.
\label{eq:108}
\eeq
Since during the De Sitter era the scale factor gets larger and larger, 
the term proportional to B in the above equation becomes more and more 
negligible. Thus, the perturbation $h_k$ remains (almost) constant while outside 
the Hubble radius. So, the longer the tensorial-perturbation mode remains outside the 
Hubble radius, the more it gets amplified (with respect to the case it stayed 
always inside the Hubble radius). During the RD era, the mode is again 
under the barrier, and the solution is 
\beq
\psi_k = \alpha_k\,e^{-i k \eta}+ \beta_k\,e^{+i k \eta}\,,
\label{eq:109}
\eeq
and contains both positive and negative modes. Even normalizing the initial 
state to positive frequency mode, the final state is a mixture of 
positive and frequency modes. In a quantum field theory language, such a mixing 
represents a process of pair production from vacuum. 
The coefficients $\alpha_k$ and $\beta_k$ are called Bogoliubov coefficients 
and can be obtained imposing the continuity of the tensorial perturbation and 
its first time derivative at the transition 
between cosmological phases. 

The intensity of the stochastic GW background can be 
expressed in terms of the number of gravitons per cell 
of the phase space $n_f$ with $f=|\mathbf{k}|/(2\pi)$. 
For an isotropic stochastic GW background $\rho_{\rm GW} = 
2 \int d^3k/(2\pi)^3\,(k\,n_k)$, thus 
\beq
\Omega_{\rm GW}(f) = \frac{1}{\rho_c}\,16\pi^2\,n_f\,f^4\,.
\label{spectbog}
\eeq
where $n_f = |\beta_f|^2$. 

The stochastic GW spectrum produced during 
{\it slow-roll} inflation, decreases as $f^{-2}$ in the frequency window 
$10^{-18}\mbox{--}10^{-16}$\,Hz, and then slowly decreases up to a frequency 
corresponding to modes whose physical frequency becomes less than the maximum 
causal distance during the reheating phase (which is of order of a few GHz). 
Its magnitude depends on  both the value of the Hubble parameter during inflation 
and a number of features characterizing the Universe evolution after the 
inflationary era --- for example tensor anisotropic stress due to free-streaming 
relativistic particles, equations of state, etc.~\cite{SWj,new}. 
An upper bound on the spectrum can be obtained from the measurement of the 
quadrupole anisotropy of the CMB, as seen in Sec.~\ref{sec8.1}. 
Since for a generic slow-roll inflationary model the spectrum is (weakly) 
decreasing with frequency this implies an upper bound $h_0^2\,\Omega_{\rm GW}\sim 5\times 10^{-16}$ 
at frequencies around $f\sim 100$\, Hz , where ground-based detectors reach the best 
sensitivity. 
The spectrum predicted by the class of single-field inflationary models is then 
too low to be observed by ground-based and also space-based detectors. It is therefore evident 
that a background satisfying the bound imposed by the observed amount of CMB 
anisotropies at large scales could be detected by GW interferometers 
provided that its spectrum grows significantly with frequency. This could happen 
in bouncing Universe cosmologies, such as pre--big-bang scenario~\cite{PBB,GWPBB}, 
the ekpyrotic models~\cite{GWEKP} (although the amplitude of the GW spectrum is too 
low to be observable) and in quintessential inflation~\cite{GWQuint}.

Finally, a stochastic GW background can be detected by correlating 
two GW interferometers~\cite{GWdet}. The upper 
limit on a flat-spectrum set by the LIGO Scientific 
Collaboration is $h_0^2\Omega_{\rm GW} \simeq 6.5 \times 10^{-5}$ in the frequency 
band $70 \mbox{--} 156$ Hz~\cite{lscstoch}. For frequency-independent spectra, 
the expected upper limit for the current LIGO configuration is 
$h_0^2\,\Omega_{\rm GW}< 5\times 10^{-6}$, while for advanced LIGO project is 
$h_0^2\,\Omega_{\rm GW}\sim 8\times 10^{-9}$.

\section{Acknowledgments}
I wish to thank the organizers of the Les Houches School for having invited me to 
such a pleasant and stimulating school and all the students for their interesting questions.
I acknowledge support from NSF grant PHY-0603762 and from the 
Alfred Sloan Foundation. I wish to thank also Francis Bernardeau and Christophe Grojean 
for their patience.

\end{document}